\documentclass[fleqn,usenatbib]{mnras}
\usepackage{newtxtext,newtxmath}
\pdfoutput=1

\usepackage[T1]{fontenc}

\DeclareRobustCommand{\VAN}[3]{#2}
\let\VANthebibliography\thebibliography
\def\thebibliography{\DeclareRobustCommand{\VAN}[3]{##3}\VANthebibliography}

\usepackage{graphicx}	
\usepackage{amsmath}	

\title[Ionised extraplanar gas in NGC 3982 and NGC 4152]{A Kinematic Analysis of Ionised Extraplanar Gas in the Spiral Galaxies NGC 3982 and NGC 4152}

\author[A. Li et al.]{
Anqi Li,$^{1}$\thanks{E-mail: li@astro.rug.nl}
Antonino Marasco,$^{2}$
Filippo Fraternali,$^{1}$
Scott Trager$^{1}$
and Marc A. W. Verheijen$^{1}$
\\
$^{1}$Kapteyn Astronomical Institute, University of Groningen, Landleven 12, 9747 AD Groningen, The Netherlands\\
$^{2}$INAF–Osservatorio Astrofisico di Arcetri, Largo E. Fermi 5, I-50157, Firenze, Italy\\
}

\date{Accepted April 6, 2021; Received February 15, 2021; in original form November 9, 2020 }

\pubyear{2021}

\begin{document}
\label{firstpage}
\pagerange{\pageref{firstpage}--\pageref{lastpage}}
\maketitle

\begin{abstract}
We present a kinematic study of ionised extraplanar gas in two low-inclination late-type galaxies (NGC~3982 and NGC~4152) using integral field spectroscopy data from the DiskMass H$\alpha$ sample. We first isolate the extraplanar gas emission by masking the H$\alpha$ flux from the regularly rotating disc. The extraplanar gas emission is then modelled in the three-dimensional position-velocity domain using a parametric model described by three structural and four kinematic parameters. Best-fit values for the model are determined via a Bayesian MCMC approach. The reliability and accuracy of our modelling method are carefully determined via tests using mock data. We detect ionised extraplanar gas in both galaxies, with scale heights $0.83^{+0.27}_{-0.40}\,\mathrm{kpc}$ (NGC~3982) and $1.87^{+0.43}_{-0.56}\,\mathrm{kpc}$ (NGC~4152) and flux fraction between the extraplanar gas and the regularly rotating gas within the disc of 27\% and 15\% respectively, consistent with previous determinations in other systems. We find lagging rotation of the ionized extraplanar gas in both galaxies, with vertical rotational gradients $-22.24^{+6.60}_{-13.13} \,\mathrm{km\,s^{-1}\,kpc^{-1}}$ and $-11.18^{+3.49}_{-4.06}\,\mathrm{km\,s^{-1}\,kpc^{-1}}$, respectively, and weak evidence for vertical and radial inflow in both galaxies. The above results are similar to the kinematics of the neutral extraplanar gas found in several galaxies, though this is the first time that 3D kinematic modelling of ionised extraplanar gas has been carried out. Our results are broadly consistent with a galactic fountain origin combined with gas accretion. However, a dynamical model is required to better understand the formation of ionised extraplanar gas. 
\end{abstract}

\begin{keywords}
galaxies: haloes -- galaxies: ISM -- galaxies: evolution -- ISM: structure -- ISM: kinematics and dynamics
\end{keywords}



\section{Introduction}
\label{sec:introduction}

Spiral galaxies are surrounded by multiphase gas layers extending up to several kpc from their disc planes (see \citealt{Reynolds73,Lockman02,Marasco11} for observations of the Milky Way's gas layers and \citealt{Dettmar90,Rand90,Swaters97,Oosterloo07} for observations of other galaxies), which are detected both in emission (\ion{H}{i} and H$\alpha$) and in low- and high-ions absorption against bright background sources \citep{Lehner11,Zheng17}. These gas layers,  known as extraplanar gas (EPG), are located at the disc-halo interface where gas accretion and stellar feedback take place. Studying their properties gives us fundamental insight into  the gas cycle between galactic discs and the circumgalactic medium (CGM), which in turn is key to understanding galaxy formation and evolution mechanisms.

Neutral hydrogen gas (\ion{H}{i}) is a primary tracer of EPG, and extensive observations of neutral EPG have been conducted. Studies find that neutral EPG is ubiquitous in late-type galaxies, with \ion{H}{i} mass fractions relative to the underlying neutral gas mass in the disc varying from 10\% to 30\% (e.g.\ \citealt{Oosterloo07,Marasco19}). Observations of nearby spiral galaxies with different inclination angles (nearly face-on: \citealt{Boomsma08}; intermediate inclination: \citealt{Fraternali01,Hess09,Marasco19}; and edge-on: \citealt{Oosterloo07,Rand08,Zschaechner11}) provide a 3D view of the structure and kinematics of neutral EPG in disc galaxies.  Neutral EPG has a scale height around 1--3\,kpc, and its velocity is characterised by differential rotation similar to the disc, but lagging behind the disc gas with a vertical velocity gradient of typically $-10$ to $-20\,\mathrm{km\,s^{-1}\,kpc^{-1}}$ \citep{Fraternali01, Fraternali05, Oosterloo07,Zschaechner11, Marasco19}. Non-circular motions like radial inflow and vertical flows  are also detected in the EPG of some galaxies  (e.g.\ \citealt{Fraternali02, Barbieri05, Marasco19}).

The presence of EPG in the Milky Way has been known for several decades. It was first detected in the form of extended \ion{H}{i} complexes with velocities that deviate strongly from the Galactic rotation \citep{Muller63,Oort66}. Clouds with deviation velocities larger than 90$\,\mathrm{km\,s^{-1}}$ are called high-velocity clouds (HVCs) \citep{Wakker97,Westmeier07}. HVCs usually have sub-solar metallicities and are located at several $\mathrm{kpc}$ from the Galactic plane \citep{Wakkers01}. Although first detected in HVCs, most extraplanar \ion{H}{i} clouds in the Milky Way are at intermediate velocities (referred to as IVCs), which have near-solar metallicities and are within 1--2$\,\mathrm{kpc}$ of the Sun \citep{Wakkers01}. A kinematic modelling study of the global EPG emission in the Milky Way shows that the EPG in the Milky Way has a velocity gradient around $-15\,\mathrm{km\,s^{-1}\,kpc^{-1}}$ and inflow in the radial and vertical directions \citep{Marasco11}. This model shows that the ``classical" IVCs are simply the local (Solar Neighbourhood) manifestation of a widespread (across the whole Galaxy) EPG component.

Ionised EPG has also been detected in the Milky Way and several other galaxies (e.g.\ \citealt{Dettmar90,Rand90,Fraternali04}), which has been summarised in the review of \cite{Haffner09}. Ionised EPG in the Milky Way is detected in multiple tracers: in H$\alpha$ emission-line observations \citep{Putman03,Haffner03}, in warm absorbers (\ion{Si}{iii}, \ion{Si}{iv}, \ion{C}{ii}, \ion{C}{iv}; \citealt{Lehner12,Zheng17}) and in \ion{O}{iv} absorbers \citep{Sembach03,Savage03}. 
Photometric observations of large galaxy samples have shown that ionised EPG is common in late-type galaxies \citep{Rossa03a,Rossa03b,Miller03a}. Spectroscopic studies \citep{Miller03b,Fraternali04,Heald05} and IFU surveys \citep{Jones17,Ho16,Bizyaev17,Levy19} have revealed the following properties of ionised EPG: (1) ionised EPG has a scale height of several kpc \citep{Rossa03a,Miller03a,Levy19} and (2) ionised EPG also shows a lagging rotation, with typical velocity gradients around $-10$ to $-20\,\mathrm{km\,s^{-1}\,kpc^{-1}}$ \citep{Kamphuis07, Miller03b, Fraternali04, Heald06a,Bizyaev17}. Most ionised EPG studies are limited to edge-on galaxies, and thus vertical motions are not accessible. Only two observations have been conducted on intermediate-inclination galaxies, both of which suggest a galaxy-scale inflow \citep{Fraternali04,Zheng17}. 

The similarity between neutral and ionised EPG kinematics suggests that these two phases might have the same origin. \citet{Rossa03a} show that galaxies with larger star-formation rates tend to possess more prominent ionised EPG (also seen in \citealt{Jones17}). Previous studies find that the distribution of the EPG is often correlated with the locations of the \ion{H}{ii} regions in the disc \citep{Miller03a,Boomsma08}.  The ionisation source for ionised EPG is most likely to be photons leaking from star-forming regions \citep{Levy19}. Thus, a possible scenario is that the ionised EPG and neutral EPG are counterparts: part of neutral EPG is photoionised by star formation activity and turns into ionised EPG.

The origin of EPG is a long-standing issue, and two different mechanisms have been proposed. One is an internal mechanism: a galactic fountain powered by star formation activities like supernova explosions or stellar winds ejects gas above the galactic disc to form the EPG (e.g.\ \citealt{Shapiro76,Bregman80}). The other is an external mechanism: accretion of gas from the CGM \citep{Binney05,Kaufmann06}. A more plausible explanation of the origin of EPG involves both mechanisms (discussed below).

A galactic fountain is the best-studied mechanism for EPG formation and maintenance. In general, \ion{H}{i} gas in the disc is partly ionised by photons from star-forming regions and pushed above the disc by supernova explosions or stellar winds. The gas then travels through the halo within the gravitational potential of the galaxy, cools down and eventually falls back to the disc. A strong prediction of such a model is that the EPG kinematics should follow the galaxy rotation, which is confirmed by the observations. Simple ballistic models for the galactic fountains reproduce many of the observed properties of the EPGs (e.g.\ \citealt{Fraternali06}). Nevertheless, an external origin of EPG cannot be ruled out, as a pure galactic fountain model tends to predict a radial outflow and underestimate the velocity gradient of the EPG \citep{Heald06a,Fraternali06}. To solve these issues, an external mechanism that can alter the angular momentum of the EPG is required. A possible scenario is fountain-driven accretion, where interaction with the fountain's neutral clouds decreases the cooling time of the hot coronal gas to a time scale shorter than a cloud’s orbit time, and thus part of the hot gas is condensed and accreted onto to the disc \citep{Fraternali17}. Models based on this scenario explain the EPG kinematics observed in external galaxies \citep{Fraternali08} and the phase-space distribution of the cold (neutral) and warm (ionised) EPG in the Milky Way \citep{Marasco12,Marasco13,Fraternali13}. 

A comprehensive 3D view of the multi-phase EPG kinematics is needed to disentangle the origin of the EPG. \citet{Marasco19} studied neutral EPG in a sample of 15 galaxies at intermediate inclinations using the HALOGAS survey \citep{Heald11}. They modelled the EPG and characterised its structural and kinematic properties, focusing in particular on the spatial distribution, vertical and radial velocities, rotational gradient, and velocity dispersion, providing the first systematic and homogeneous study of neutral EPG in nearby disc galaxies.

As for ionised EPG, however, nearly all samples are composed of edge-on galaxies, which limits or even prevents analysis of non-circular (especially vertical) motions of the EPG. We therefore have chosen to study the ionised EPG of a pair of intermediate-inclination galaxies in this paper. In Sec.~\ref{sec:data} we introduce the DiskMass sample from which we select our galaxies. The strategy used for modelling ionised EPG, adapted from \citet{Marasco19}, is described in Sec.~\ref{sec:method} and tested with mock galaxies in Sec.~\ref{sec:mock}. We present our analysis and results for ionised EPG in a pair of DiskMass sample galaxies in Sec.~\ref{sec:results} and discuss the results in Sec.~\ref{sec:discussion}. We summarise our results in Sec.~\ref{sec:summarise}.

\section{Data Reduction}
\label{sec:data}

\subsection[The Halpha sample of the DiskMass Survey]{The H$\alpha$ sample of the DiskMass Survey}
\label{sec:diskmass}
In this paper, we use the DiskMass survey (DMS) H$\alpha$ sample (\citealt{Bershady10}; Swaters et al., in prep.) to study the existence and kinematic properties of ionised extraplanar gas (EPG). This sample contains integral field spectroscopy of 138 late-type galaxies in the DMS parent sample \citep{Bershady10} observed with the SparsePak integral field unit on the WIYN telescope \citep{Bershady05}. SparsePak is a sparsely distributed fibre optic bundle containing 82 fibres, including seven sky fibres. Each fibre has a diameter of $4\farcs7$ and is separated by $5\farcs6$, and the fibres are sparsely distributed on the focal plane. All 138 galaxies have low or intermediate inclination angles and diameters comparable to the field-of-view of SparsePak ($1\arcmin < D_{25} < 3\farcm5$).

 Each galaxy in the DMS H$\alpha$ sample has been observed with three pointings of different offsets.  One pointing is centred at the galaxy centre, a second pointing is $5\farcs6$ to the south, and a third pointing is $4\farcs9$ to the west and $2\farcs8$ to the north. Since the distribution of fibres follows a hexagonal pattern \citep{Bershady05}, the SparsePak field-of-view can be almost fully sampled with this 3-pointing strategy. 

The SparsePak signal is fed into the Bench Spectrograph, a spectrograph optimised for use with fibre optic bundles. For the DMS H$\alpha$ sample, the 316 $\mathrm{lines\,mm^{-1}}$ echelle grating in order 8 was used, which covered a wavelength range of 6475--6880{\,\AA}, with a dispersion of $0.195\,\mathrm{\AA\,pix^{-1}}$ and an instrumental FWHM of $30\,\mathrm{km\,s^{-1}}$.
 
The spectral data we use in this paper have been reduced in the following manner (Swaters et al, in prep). The raw data have been over-scan corrected and trimmed, and a bias level has been subtracted. Cosmic rays have been removed using the method introduced in \citet{Andersen06}. The IRAF\footnote{Developed by USA National Optical Astronomical Observatories (NOAO): \url{https://iraf-community.github.io/}} task \textit{dohydra} has been used to flat-field, wavelength-calibrate, and extract the spectra. Sky and (stellar) continuum have then been removed. The data product for each pointing is a FITS file in row-stacked spectra (RSS) format that contains 82 emission-line spectra with a wavelength range of 6475--6880 {\AA} and a wavelength sampling of $0.2\,\mathrm{\AA\,pix^{-1}}$, which corresponds to $9.14\,\mathrm{km\,s^{-1}}$ around the H$\alpha$ line.

\subsection{The subsample in this paper}
In this work we focus on two galaxies from the DMS sample, NGC~3982 and NGC~4152. We select these two galaxies because they have sufficient S/N to reveal the relatively weak EPG emission, and their structures are well-resolved under the SparsePak spatial resolution. Besides, their position--velocity (pv) diagrams along their major axes have emission features called `beard' patterns \citep{Schaap00,Fraternali01} that deviate from the bulk rotation and extended towards the systemic velocity. These two galaxies are the best cases in the DMS to study the EPG. This does not mean that other galaxies in the DMS sample do not have EPG, but rather they do not have high-enough S/N or spatial resolution to allow us comment on the incidence of EPG.

\begin{figure*}
\centering
\includegraphics[scale=0.5]{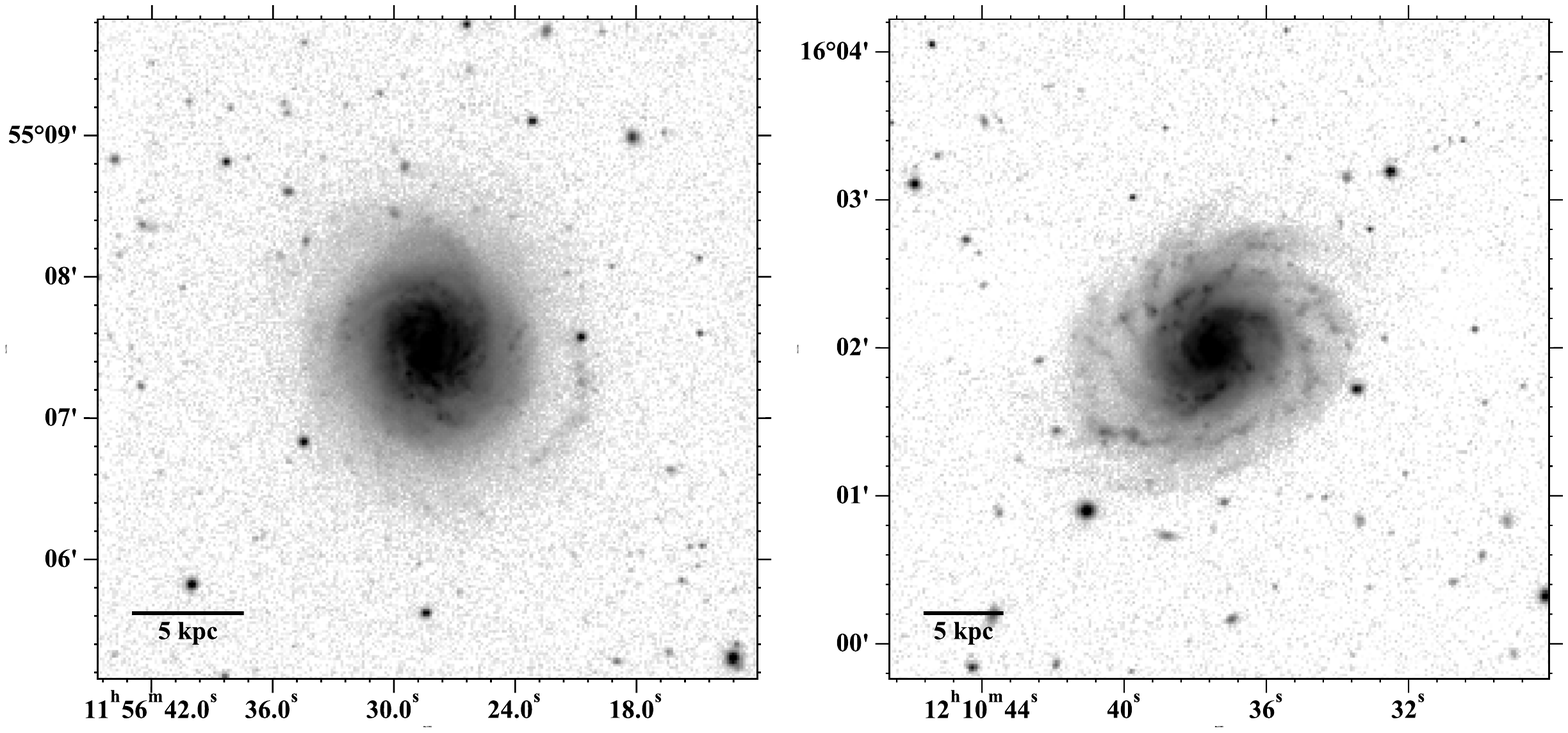}
\caption{The $r$-band images from SDSS DR4 \citep{Baillard11}. Left: NGC~3982. Right: NGC~4152.}
\label{fig:image}
\end{figure*}

NGC3982 is a late-type Seyfert 2 galaxy \citep{Cardamone07} with a relatively low inclination angle ($26\fdg2$) and a distance of $21.6\,\mathrm{Mpc}$ \citep{Tully13}. It is a high surface brightness member of the Ursa Major cluster, with on-going star-forming activity \citep{Martinsson13}, making it an ideal target for the kinematic study of EPG. NGC~4152 is a late-type galaxy with on-going star formation. It lies at a distance of $30.6\,\mathrm{Mpc}$ \citep{Tully09}, with an inclination angle of $38\fdg1$. Figure~\ref{fig:image} shows the $r$-band images of NGC~3982 and NGC~4152. Their physical properties are listed in Table~\ref{tab:galaxies}. 

\begin{center}
\begin{table*}
\begin{tabular}{lrrrrrrr}
\hline
\multicolumn{1}{c}{Galaxy Name}&\multicolumn{1}{c}{PA}&\multicolumn{1}{c}{INCL}&\multicolumn{1}{c}{Distance}&\multicolumn{1}{c}{$\mathrm{V_{sys}}$}&\multicolumn{1}{c}{Hubble Type}&\multicolumn{1}{c}{Absolute Mag }&\multicolumn{1}{c}{$\mathrm{V_{flat}}$} \\
\multicolumn{1}{c}{(1)} & \multicolumn{1}{c}{(2)} & \multicolumn{1}{c}{(3)} & \multicolumn{1}{c}{(4)} & \multicolumn{1}{c}{(5)} & \multicolumn{1}{c}{(6)} & \multicolumn{1}{c}{(7)} & \multicolumn{1}{c}{(8)}\\
 \hline \hline
 NGC~3982&$190.0\pm3.6$&$26.2\pm5.2$&21.6$^a$&$1113.6\pm1.8$&Sb$^d$ &$-21.14\pm0.11^e$ &$221.9\pm34.3$\\
 NGC~4152&$324.4\pm1.9$&$38.1\pm3.5$&30.6$^b$&$2174.6\pm1.2$&Sc$^d$ &$-20.68\pm0.32^f$&$134.5\pm15.9$\\
\hline
\end{tabular}

\caption{Galaxy properties. Columns: (1) Galaxy name.  (2)--(3): Position-angle and inclination derived from this work with the Groningen imaging processing system (GIPSY, \citealt{Vanderhulst92}) task \textit{rotcur}, in units of degree. (4) Distance of galaxies from literature, in units of Mpc: $^a$\citet{Tully13}; $^b$\citet{Tully09}. (5) Systematic velocities derived from this work with GIPSY task \textit{rotcur}, in units of $\mathrm{km\,s^{-1}}$. (6) Hubble type from literature: $^d$\citet{deVaucouleurs91}. (7) Absolute magnitude in visible band: $^e$\citet{Cardamone07}; $^f$\citet{Kim14}. (8) Flat rotation velocity derived from this work with $\mathrm{^{3D}}$Barolo, in units of $\mathrm{km\,s^{-1}}$. The quoted error-bars account for the uncertainties in the inclinations.}
\label{tab:galaxies}
\end{table*}
\end{center}

\begin{figure*}
\centering
\includegraphics[scale=0.5]{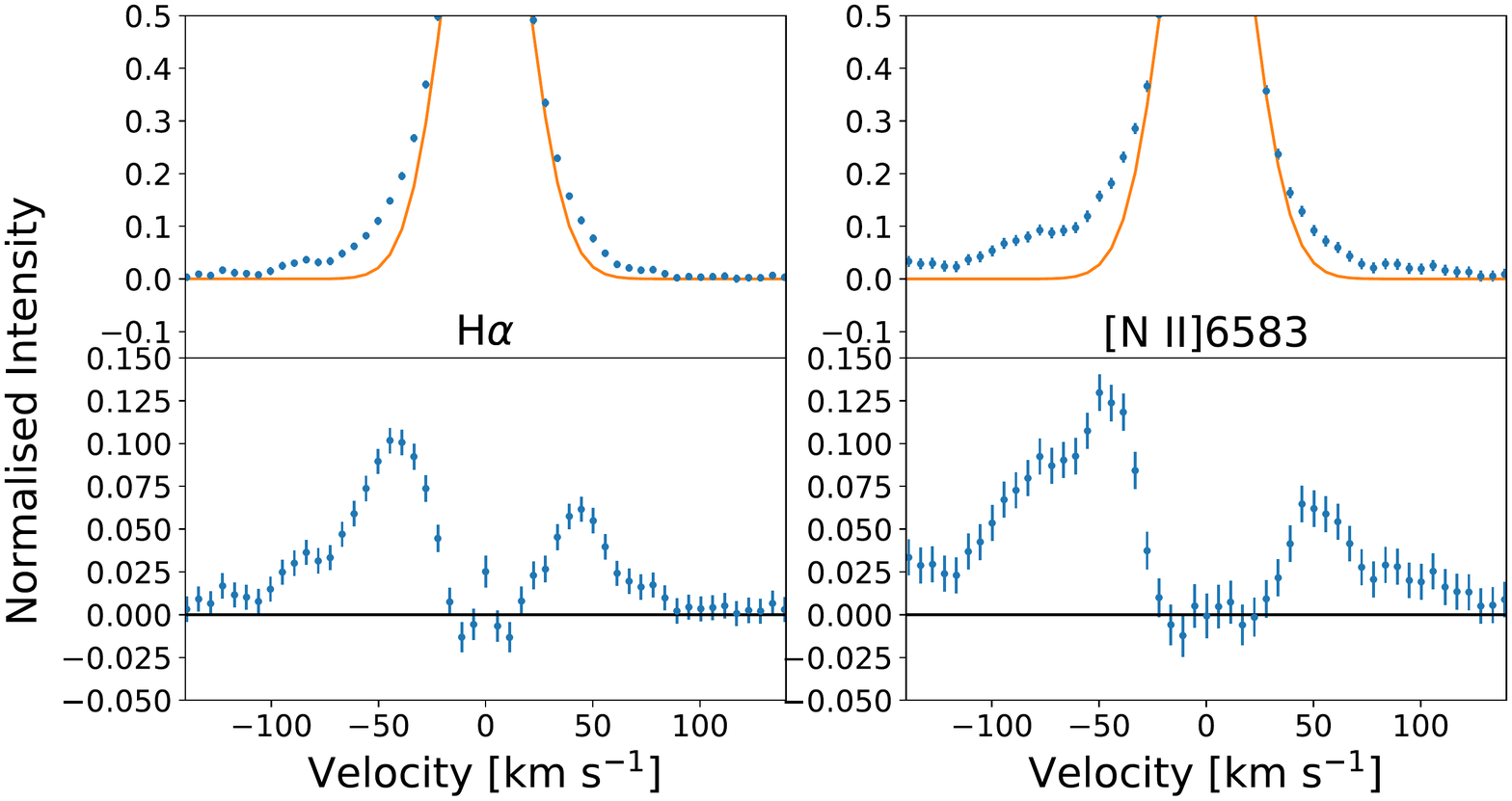}

\includegraphics[scale=0.5]{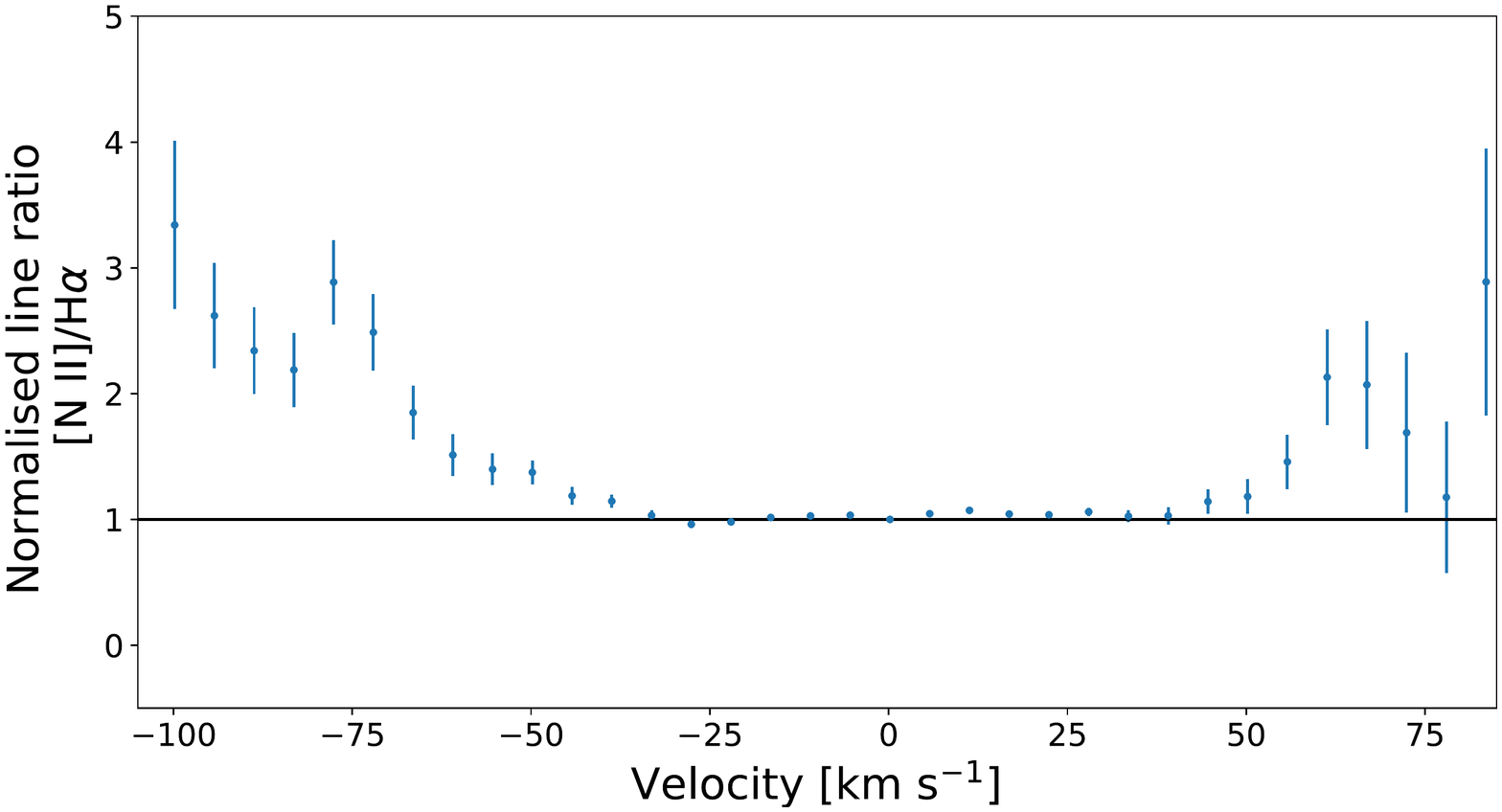}
\caption{Top-left panel: the stacked H$\alpha$ spectrum within $\pm$\,30$\degr$ of the major axis of NGC 3982. The scale is compressed to better show the tails of emission. The continuous orange curve indicates the Gaussian fit. Residuals are shown below. Top-right panel: as in the top-left panel, but for the [\ion{N}{ii}]\,6583 line. Bottom panel: [\ion{N}{ii}]\,6583/H$\alpha$ ratio versus velocity difference from the rotation of the disc, normalised to the ratio measured at $v=0\,\mathrm{km\,s^{-1}}$.}
\label{fig:ratio}
\end{figure*}

To get a first glimpse of the EPG in the data, we start by stacking spectra of NGC~3982 before implementing the 3D-modelling analysis. We fit a Gaussian function to every {H$\alpha$} velocity profile in NGC~3982 that is within {$\pm$30\degr} of the major axis and then use the fitted intensities to normalise each profile. All normalised profiles are shifted using the fitted central velocity of the Gaussian (spectra on the receding side are flipped). In this way all velocity profiles are centred at zero velocity while positive (negative) values represent rotation faster (slower) than the disc. We then stack those profiles together and get a stacked H$\alpha$ spectrum for NGC~3982. The wavelength range of our spectra data also covers the [\ion{N}{ii}] doublet (6583\,{\AA} and 6548\,\AA), although the S/N of the [\ion{N}{ii}] doublet does not allow us to detect EPG in most single velocity profiles. Using a similar method, we also derive a stacked [\ion{N}{ii}]\,6583 spectrum. The stacked spectra overlaid by Gaussian fits (based on the upper 40\% of the profiles) are shown in the top panel of Fig.~\ref{fig:ratio}, also shown are the residuals of the Gaussian fit. It is clear that both H$\alpha$ and [\ion{N}{ii}]\,6583 profiles show wings that cannot be fitted by a single Gaussian, which are very likely due to EPG. Note that NGC~3982 is a Seyfert galaxy, the stacked spectra might be contaminated by active galactic nucleus (AGN) feedback. We have therefore also tried masking the central 1\,kpc region and verified that the pattern of the wings does not change with respect to that shown in Fig~\ref{fig:ratio}. As mentioned, the kinematics of the EPG are characterised both by a lag in rotation with respect to the disc gas and non-circular (vertical and radial) motions. The fact that the wing at the slower-rotation side is stronger is what we would expect from EPG, since the lag only contributes to the slower-rotation side while vertical motion results in wings on both sides. We also look into the [\ion{N}{ii}]\,6583/H$\alpha$ ratio of the stacked spectra, which is shown in the bottom panel of Fig.~\ref{fig:ratio} and normalised at zero velocity. The ratio increases for gas at velocities that depart from the disc rotation (peak of the Gaussian) as expected, since EPG usually shows a higher [\ion{N}{ii}]\,6583/H$\alpha$ ratio compared to  gas in the disc \citep{Hoopes99}, similar to what \citet{Fraternali04} found for ionised EPG in NGC~2403.

\begin{figure*}
\centering
\includegraphics[scale=0.5]{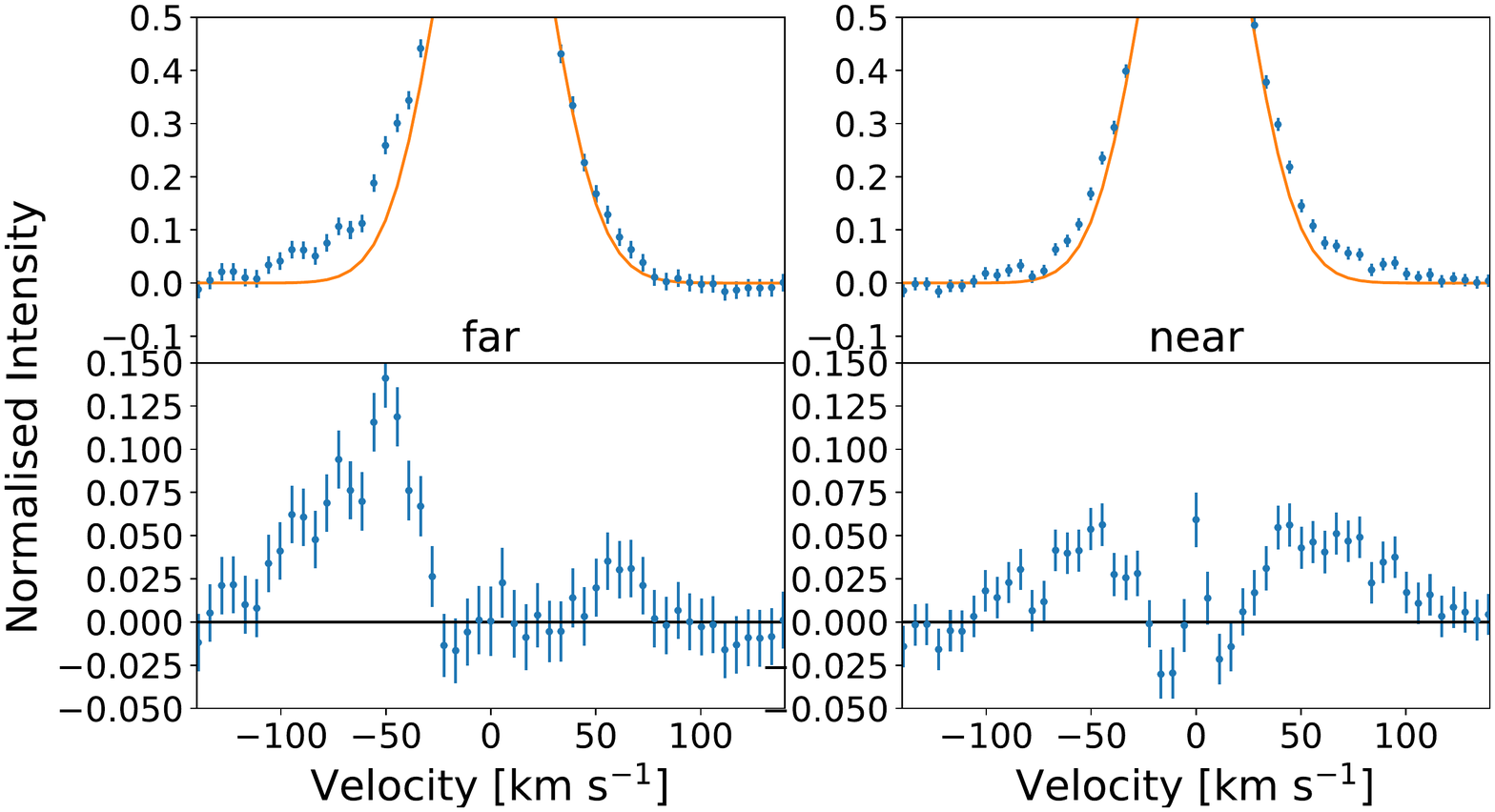}
\caption{Left panel: The stacked H$\alpha$ spectrum within $\pm$\,30$\degr$ of the minor axis at the far side of NGC~3982. The scale is compressed to better show the tails of emission. The continuous orange curve indicates the Gaussian fit. Residuals are shown below. Right panel: as in the left panel, but for the near side of NGC~3982.}
\label{fig:minor-stack}
\end{figure*}

The stacked spectra along the major axis of NGC~3982 do not give us information about radial motions, which are actually often detected in EPG, as mentioned in Sec.~\ref{sec:introduction}. Radial motions are visible in spectra along the minor axis, where inflow (outflow) will cause a blue-shifted (red-shifted) wing at the far side of the galaxy and a red-shifted (blue-shifted) wing at the near side of the galaxy. We therefore also derive stacked H$\alpha$ spectra within {$\pm$30\degr} of the minor axis using similar method as described above, but stacking the far side and the near side separately, as shown in Fig.~\ref{fig:minor-stack}. The blue-shifted wing dominants in the far-side stacked spectrum while the red-shifted wing is relatively more prominent in the near-side stacked spectrum, suggesting a radial inflow for EPG in NGC~3982, consistent with studies of other galaxies (see Sec.~\ref{sec:introduction}).

The above analysis on the stacked spectra strongly suggests that ionised EPG is present in NGC~3982. In the following we further study the kinematics of ionised EPG by 3D modelling.

\subsection[Building Halpha datacubes]{Building H$\alpha$ datacubes}
\label{sec:datacubes}

The 3D modelling code we use to study EPG is implemented on datacubes \citep{Marasco19}, but the DMS H$\alpha$ data is in RSS format, as described above. We take the following steps to reshape the spectra of NGC~3982 and NGC~4152 into datacubes.

First we build a model FITS file for each galaxy with the Groningen imaging processing system (GIPSY, \citealt{Vanderhulst92}) task \textit{create}. This FITS file is a $300\times 300$ pixel 2D image with a pixel size of $0\farcs5$. As mentioned above in Sec.~\ref{sec:diskmass}, the distribution of SparsePak fibres follows a hexagonal pattern. We need an over-sampled FITS file to allocate fibres with high accuracy, and therefore the grid size is much smaller than the $4\farcs7$ fibre diameter of SparsePak. The FITS file is centred at the galaxy centre. 

Next we convert the wavelengths of the spectra into optical velocities with respect to the H$\alpha$ line and use the IRAF task \textit{rvcorrect} to derive the heliocentric velocity (all velocities in this paper are heliocentric velocities or velocities relative to the galaxy centre). We have spatial offsets for each of the 75 fibres (ignoring seven sky fibres) in each pointing, as well as the spectrum at each pointing. For every point along this velocity grid, we then use the GIPSY task \textit{gds2text} to fill the model FITS image with intensities at that particular velocity of all $75\times3$ spectra from the three pointings. Finally, we have an image for each velocity along the velocity axis, which is equivalent to a slice of a datacube along the velocity axis.

Then we use GIPSY task \textit{create} again to build a 3D model FITS file with a size of $300\times300\times81$ pixels. Its spatial grid size ($0\farcs5\,\mathrm{pix^{-1}}$) is that of the above-mentioned 2D model FITS file while the third axis is velocity, centred at the systematic velocity of the galaxy with a velocity interval of $9.14\,\mathrm{km\,s^{-1}}$ (equivalent at H$\alpha$ to the underlying spectrum's wavelength interval of $0.2\,\mathrm{\AA\,pix^{-1}}$) and a velocity resolution of $30\,\mathrm{km\,s^{-1}}$, corresponding to spectral resolution of 10049. With 81 channels we cover a velocity range of $\pm365.44\,\mathrm{km\,s^{-1}}$ around the systemic velocity of the galaxy, sufficient to capture the kinematics of the disc gas and EPG. We then pile the image slices from the above step into this 3D model FITS file. 

Finally, we have a datacube for each galaxy in the subsample that contains spectra from all three pointings. Note that because our field-of-view was not fully sampled, there are blank regions in the datacube. We are aware that the above method is different from the usual way of building IFU datacubes where interpolation is done between adjacent pixels and blank regions are filled. We choose this method under the consideration that the study of EPG kinematics prefers a small beam-smearing effect. However, the beam-smearing effect with our relatively large fibre size ($4\farcs7$, corresponding to $\sim$\,0.6\,kpc in our sample) is already non-negligible and would increase further if we interpolated between different fibres. Thus we build the datacube as appropriate to the original observation, at the cost of blank regions in the resulting datacube where no data were available.
\section{Modelling the extraplanar gas}
\label{sec:method}

We use the EPG modelling code written by \citet{Marasco19} to study the EPG in NGC~3982 and NGC~4152, which was originally written with the aim of studying the neutral \ion{H}{i} EPG in the HALOGAS survey \citep{Marasco19} but in principle is also capable of characterising structure and kinematics of generic line-emitting medium outside the disc.    

The spatial resolution in the HALOGAS sample is $30\arcsec$, corresponding to $2.6\,\mathrm{kpc}$ for NGC~4414, which is the furthest galaxy in the HALOGAS sample for which the code manages to model the EPG \citep{Marasco19}. As a comparison, the spatial resolution in our datacube is {$4\farcs7$}, corresponding to $0.41\,\mathrm{kpc}$ for NGC~3982 and $0.60\,\mathrm{kpc}$ for NGC~4152.  The typical velocity resolution in the HALOGAS sample is $\mathrm{FWHM}\,\sim\,8\,\mathrm{km\,s^{-1}}$ \citep{Heald11,Marasco19} while in our datacube the FWHM is around $30\,\mathrm{km\,s^{-1}}$ . Although not comparable with the velocity resolution in the HALOGAS sample, we argue that a FWHM of $30\,\mathrm{km\,s^{-1}}$ is sufficient for kinematic study of ionised gas, and mock galaxy tests in Sec.~\ref{sec:mock} verify that the code can model the EPG successfully under such velocity resolution. This code (with some customisation) is therefore suitable for our datacubes. 

In general, the code first separates EPG from disc emission and then fits the EPG emission with an axisymmetric, smoothly distributed EPG model, which is characterised by three structural and four kinematic parameters. An MCMC algorithm is used to find the best-fit parameters for the data. In the following parts, we briefly introduce the algorithm of the code and the modifications we made for H$\alpha$ data; more details about the code  can be found in \citet{Marasco19}.  

\subsection{Isolating the EPG signal}
\label{sec:sep}
Separation of EPG emission from the underlying gas disc and noise in the datacube is done through two masks: an external mask, which filters out noise voxels without emission from the galaxy; and an internal mask, which filters out the disc emission. The external mask is created by spatially smoothing the datacube  by a factor of 5 (using a 2D Gaussian kernel), calculating a smoothed rms noise level, then sigma-clipping at $\mathrm{S/N}=4$, which removes the majority of noise voxels and leaves only voxels with emission.

\begin{figure*}
\centering
\includegraphics[scale=0.6]{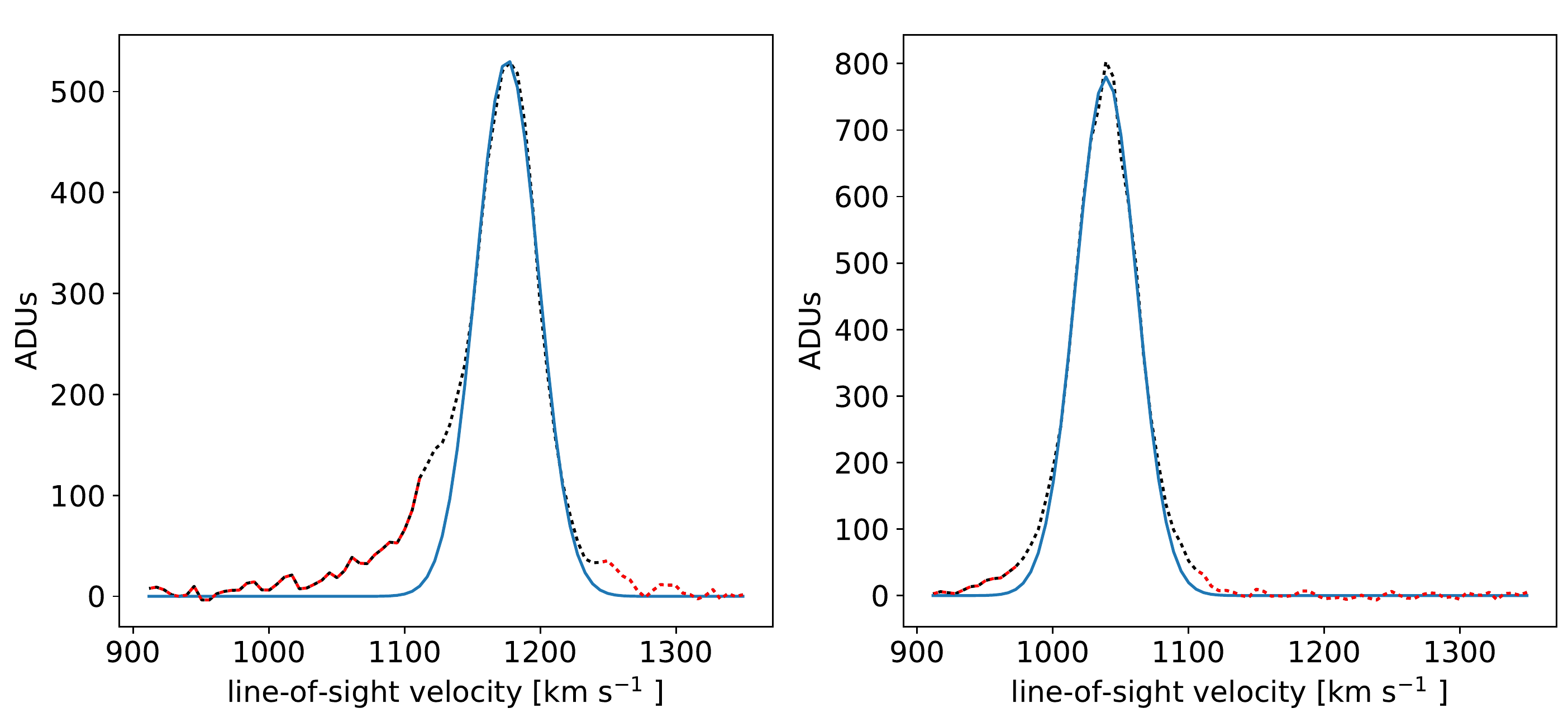}
\caption{Two velocity profiles from the datacube of NGC~3982. Dotted lines are the velocity profiles from the data, while black dots are voxels filtered out by the internal mask and red dots are EPG residual voxels used for following analysis. Blue lines are the fitted disc profiles, through which internal masks are generated.}
\label{fig:profile}
\end{figure*}

The internal mask, which separates disc emission from the EPG emission, is created using the method introduced by \citet{Fraternali02}, under the assumption that the EPG velocity deviates from the disc motion because of EPG's lagging rotation, vertical and radial motion, as well as larger velocity dispersion (see Sec.~2.2 in \citealt{Marasco19} for details).  For the velocity profile at a certain spaxel, the disc component is assumed to be described by a Gaussian profile. The lagging rotation of the EPG adds a wing to the profile at the slower rotation side, and the radial and vertical motion of the EPG contribute wings at both sides (a sketch of this can be found in Fig.~1 of \citealt{Marasco19}). Although the disc and EPG profiles are blended together, the ratio of the EPG's mass to that of the disc is relatively small, and thus it is reasonable to neglect the contribution of the EPG around the peak of the emission line profile and use this region to fit disc emission. We perform a Gaussian fit for the disc component using only the upper 40\% (in intensity) of the  line profile. We implement this method on all pixels that have passed through the external masking. By considering all the fitted Gaussian profiles together, we build a non-parametric model cube for the H$\alpha$ emission from the disc. Finally, the internal mask is generated from this synthetic-disc datacube: all voxels with disc emission higher than $N$ times the rms-noise are masked out. The mask threshold $N$ should be low enough to reject the majority of disc emission and high enough to leave as much EPG emission in the data as possible. We choose $N=2$, which was decided empirically and has been tested to be eligible for separating disc and EPG (see Sec.~2.2 of \citealt{Marasco19}). Moreover, in  Sec.~\ref{sec:ngc3982} we show that for our data the results of MCMC fit are not significantly sensitive to the mask threshold. Figure~\ref{fig:profile} shows two example velocity profiles illustrating how the internal mask works. 

\subsection{Extraplanar gas model}
\label{sec:epgmodel}
After applying external and internal masks to the datacube, we fit the remaining emission with a synthetic datacube built from a parametric model of the EPG. The density distribution of this EPG model follows the profiles introduced in \citet{Oosterloo07}, which assumes a cylindrical geometry. The surface density profile is given by \begin{equation}
\label{eq:surface}
\Sigma(R)=\Sigma_{0}\left(1+\frac{R}{R_g}\right)^{\gamma}\exp\left(-\frac{R}{R_g}\right),
\end{equation}
\noindent where $\Sigma_{0}$ is the central surface density,  $\gamma$ is an exponent regulating the density decline towards the centre, and $R_g$ is the scale length of the EPG. The density distribution along the $z$-direction is given by
\begin{equation}
\rho(z)\propto\frac{\sinh(\left|z\right|/h)}{\cosh^2(\left|z\right|/h)},
\end{equation}
\noindent where $h$ is the scale height of the EPG. $R_g$, $\gamma$, and $h$ together control the spatial distribution of the EPG model. Four other parameters characterise the EPG kinematics: the rotation velocity gradient along the $z$-direction that describes the lagging rotation of the EPG, $dv_{\phi}/dz$; the radial velocity, $v_R$; the vertical velocity, $v_z$; and the velocity dispersion, $\sigma$.  These seven parameters ($R_g$, $\gamma$, $h$, $dv_{\phi}/dz$, $v_R$, $v_z$, $\sigma$) together define an EPG model.

 The realisation of a parametric model into a datacube is described in detail in Section 2.4 of \citet{Marasco19}. However, we note that while for the \ion{H}{i}-emitting gas the line intensity is simply proportional to the gas column density at given velocity\footnote{Assuming an optically-thin regime, appropriate for low-density gas in the halo}, this is not the case for the H$\alpha$ emission, thus modifications to the previous implementation have to be made. These are described in details in Sec.~\ref{sec:modifications}. 
 
\begin{figure}
\centering
\includegraphics[scale=0.6]{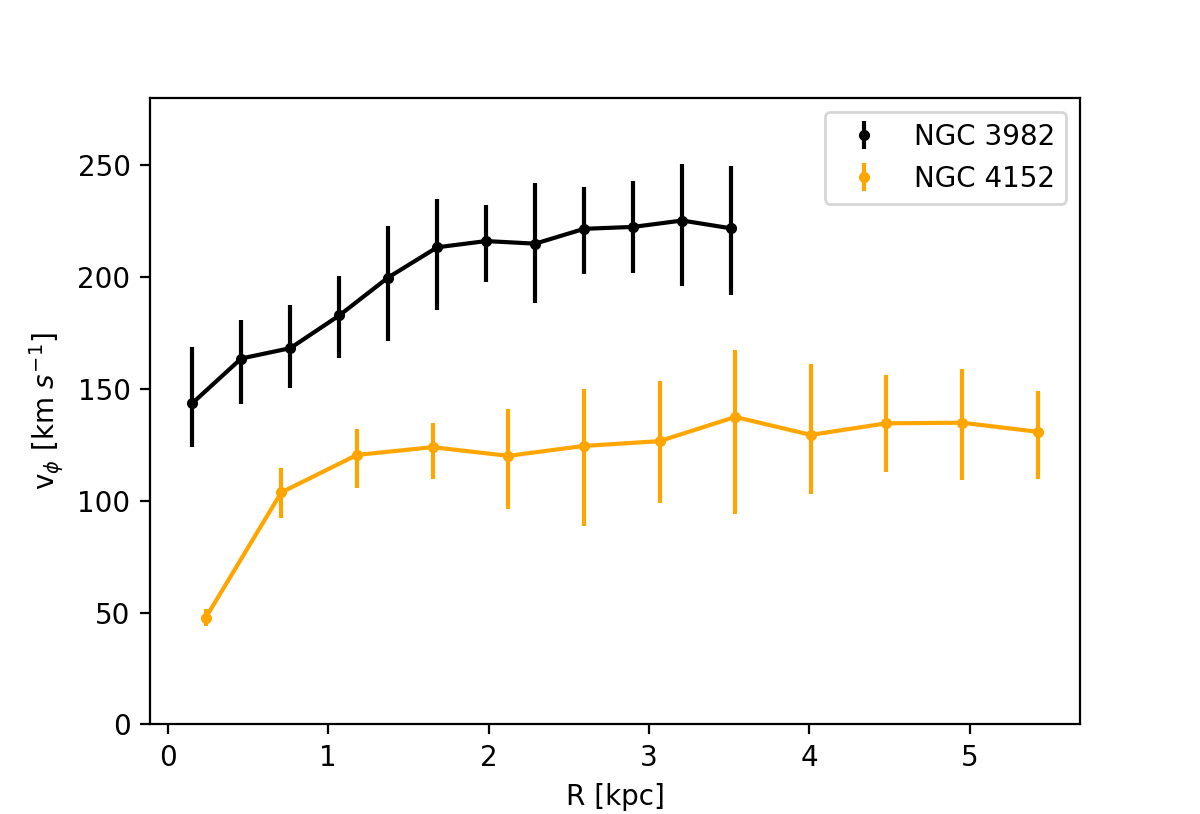}
\caption{H$\alpha$ rotation curves (with error bars) for NGC~3982 and NGC~4152, derived from $\mathrm{^{3D}}$Barolo.}
\label{fig:rot}
\end{figure}

 When building the model datacube, the distance, position angle (PA), inclination angle (INCL), and rotation curve of the galaxy are also necessary ingredients. We use distances from the literature (Table~\ref{tab:galaxies}). The rotation curve is derived from the datacube before internal masking, using the 3D-tilted-ring modelling code $\mathrm{^{3D}}$Barolo \citep{Di15}. When running $\mathrm{^{3D}}$Barolo, we fix the kinematic centre at the optical centre and use PA and INCL derived from GIPSY task \textit{rotcur} as input values, which are then slightly adjusted (by less than 1$\degr$) by $\mathrm{^{3D}}$Barolo and later implemented in the MCMC fitting described below. The inferred rotation curves for NGC~3982 and NGC~4152 are shown in Fig.~\ref{fig:rot}. It is noteworthy that the error propagated from the uncertainty on INCL is not shown in the error bars in Fig.~\ref{fig:rot}; it is around $30\,\mathrm{km\,s^{-1}}$ for NGC~3982 and $10\,\mathrm{km\,s^{-1}}$ for NGC~4152.

 \subsection{MCMC fitting}

We now estimate the values of the seven EPG model parameters using a Bayesian fitting procedure that uses MCMC for sampling the posterior.
According to Bayes' theorem, for a given data $\mathcal{D}$, the posterior probability $\mathcal{P}$ for parameter vector $\mathbf{x}$ is given by 
\begin{equation}
 \mathcal{P}(\mathbf{x}\vert\mathcal{D})\propto\mathcal{P}(\mathcal{D}|\mathbf{x})\mathcal{P}(\mathbf{x}),
\end{equation}
\noindent where $\mathcal{P}(\mathcal{D}|\mathbf{x})$ is the likelihood function and $\mathcal{P}(\mathbf{x})$ is the prior. We do not make any prior assumptions about the EPG parameters, and thus the prior for each parameter is uniform within a reasonable and wide range (see Table 3 of \citealt{Marasco19}). The likelihood function is given by

\begin{eqnarray}
\mathcal{P}(\mathcal{D}\vert\mathbf{x}) &\propto& \prod\limits_{n.voxels}^{} \exp\left({-\frac{|\mathcal{M}(\mathbf{x})-\mathcal{D}|}{\varepsilon}}\right) \nonumber \\ 
 &=&\exp\left({-\sum\limits_{n.voxels}^{}\frac{|\mathcal{M}(\mathbf{x})-\mathcal{D}|}{\varepsilon}}\right)\nonumber\\ 
 &=& \exp[-\mathcal{R}(\mathbf{x})/\varepsilon],
\end{eqnarray}

\noindent where $\mathcal{M}$ represents the model datacube built from parameter $\mathbf{x}$, $\mathcal{R}$ is the sum of the absolute residuals between the model and the data, and $\varepsilon$ is the uncertainty of the data. Since the EPG model is axisymmetric, we should take the asymmetry of the data into consideration while calculating the uncertainty. Thus we define an ``effective uncertainty'' $\varepsilon$, given by $\varepsilon=\sigma_{\delta I}\times n_{vpr}$, where $\sigma_{\delta I}$ is an estimator for the deviation of the data from pure axisymmetry, and $n_{vpr}$ is the number of voxels per resolution element  (see Eq.~6 of \citealt{Marasco19} and their Sec 2.5).
 
 We sample the posterior using \textsc{emcee}, a Python implementation of the Markov Chain Monte Carlo (MCMC) method \citep{emcee13}. During the fitting, we use 100 walkers and a chain length that varies between 1000 to 2500 steps. The walkers are initialised at a minimum, which is found by a downhill-simplex minimisation routine \citep{Nelder65}.

\subsection[Customisation for Halpha data]{Customisation for H$\alpha$ data}
\label{sec:modifications}

 As mentioned above, the EPG modelling code was written to analyse neutral EPG in the HALOGAS sample \citep{Marasco19}. The application of the code to the DMS H$\alpha$ data requires customisation, which is related to the difference in beam shapes, the conversion from gas-density to line-intensity, and the treatment of dust extinction. 
 
  The synthetic EPG datacube that we build from a parametric model to fit the EPG emission should have the same beam shape as the data. This is achieved by spatially smoothing each channel of the synthetic datacube with the beam of the data. As described in Sec.~\ref{sec:datacubes}, our datacube is built directly from the SparsePak fibre data without interpolation, and thus the data beam is equivalent to the SparsePak fibre's top-hat beam shape, i.e., uniform within each fibre's radius (2$\farcs35$) and zero outside fibres, which is contrary to the nearly Gaussian beam shape of \ion{H}{i} data (as it is in HALOGAS survey). We have therefore changed the smoothing treatment for our model datacubes accordingly.
 
 While \ion{H}{i} emission intensity is linearly proportional to the density of the emitting gas, H$\alpha$ emission depends on the square of the electron density of the extraplanar plasma \citep{Spitzer78, Collins02}. We account for this effect by simply squaring the model datacube derived by assuming linear proportionality between emission intensity and column density per velocity bin. As in \citet{Marasco19}, the cube normalisation is then determined a posteriori, by comparison with the total observed flux. We stress that this squared proportionality affects the shape of the line profiles, and as a consequence, producing a factor $\sqrt{2}$ difference between the density-weighted and flux-weighted velocity dispersions.
 
The model datacube is built in an optically-thin regime, which is a reasonable assumption for \ion{H}{i} gas. The ionised EPG gas, however, suffers from dust extinction. Due to the lack of dust extinction data for our sample, we do not introduce extinction factor into our model when fitting the EPG data. Nevertheless, mock galaxy tests show that dust extinction does not influence our analysis on ionised EPG significantly (see Sec.~\ref{sec:mock} and discussion in Sec.~\ref{sec:limitations}).

 Except for the above mentioned differences, our procedure is the same as that in \citet{Marasco19}.

\section{Testing extraplanar gas parameter estimation with mock galaxies}
\label{sec:mock}

The robustness of the code can be checked via mock data tests, where we build mock galaxies with certain parameters and see if the code can recover these parameters. Mock data tests have been conducted extensively in \citet{Marasco19} and have verified the reliability of the code. However, because of the above-mentioned differences between \ion{H}{i} data and H$\alpha$ data and the corresponding changes we make to the code, we have also conducted mock data tests.

\subsection{Building a mock galaxy }
\label{sec:buildmock}
The algorithm to build a mock galaxy is described in Appendix B of \citet{Marasco19}. We begin by choosing a distance, PA, INCL, and rotation curve for the mock galaxy. We then construct a regularly rotating thin disc model and choose an EPG model. 

We use NGC~3982 from the DMS H$\alpha$ sample as a template for building mock galaxies.  All mock galaxies (except MG7, see below) have the same distance, PA, INCL, and rotation curve as NGC~3982, using parameters inferred using the method described in Sec.~\ref{sec:epgmodel}. The disc model assumes a Gaussian profile for both vertical and surface density distribution. We set the disc scale length $R_\mathrm{d}$ to be $1.6\,\mathrm{kpc}$ (which is comparable to NGC~3982) and the disc scale height $h_\mathrm{d}$ to be $0.2\,\mathrm{kpc}$. For the disc kinematics, we assume a velocity dispersion of $\sigma_\mathrm{d} = 27\,\mathrm{km\,s^{-1}}$; we note that this is a density-weighted dispersion due to the way we build our datacube (see Sec~\ref{sec:epgmodel}), which corresponds to a flux-weighted velocity dispersion of $19.1\,\mathrm{km\,s^{-1}}$, a typical value for disc ionised gas (e.g.\ \citealt{Martinsson13}).  In Sec.~\ref{sec:epgmodel} above we have discussed how to construct an EPG model.

We then build two mock datacubes from these ingredients: a mock disc datacube and a mock EPG datacube. To produce a mock galaxy datacube, we must first decide on the ratio of the EPG flux to the disc flux. To explore the performance of the code on different flux fractions of EPG and match these with the observations -- the EPG flux fractions for NGC~3982 and NGC~4152 are 0.27 and 0.15 (Section~\ref{sec:results}) -- we choose flux fractions of 0.15, 0.20, and 0.30, covering the upper and lower fractions and an intermediate value. Once we have chosen the flux fraction, we scale the EPG and disc mock datacubes by the appropriate flux fractions and then add them to produce the mock galaxy datacube.

{Finally, we  inject noise into the mock data, assuming a constant rms noise level throughout the datacube, and quantify the corresponding ``data quality" using two parameters: [S/N]$_{\mathrm{MED}}$ and N${_\mathrm{ind}}$.  [S/N]$_{\mathrm{MED}}$ is defined as the median S/N of the voxels in the EPG cube with intensities above twice the assumed rms noise while N${_\mathrm{ind}}$ is defined as the total number of voxels in EPG residual with intensity above twice the rms-noise divided by the number of voxels per resolution element (see Eq.\,6 of \citealt{Marasco19}). As [S/N]$_{\mathrm{MED}}$ and N${_\mathrm{ind}}$ will also influence code performance,  we explore models with different values by changing the rms noise.

\subsection{Mock galaxy parameters}
\label{sec:mocktestscombination}
We first build an extreme case, MG1, where the EPG signal is strong (high S/N and flux fraction). MG2 is created by doubling the rms noise of MG1 and decreasing velocity gradient $dv/dz$ and velocity dispersion $\sigma$. MG3 is the same as MG2 except that we double the rms noise. Then we move back to the rms noise of MG2, lower the flux ratio of EPG to 0.15, flip the values for $v_R$ and $v_z$, and build MG4.

The above tests explore different S/N or EPG fractions, but it is noteworthy that in these mock galaxies the EPG has kinematics which significantly deviates from the disc (large $dv/dz$ and $h$). To further explore the parameter space, we test models with moderate S/N and EPG fraction but have EPG kinematics which deviates less from the disc. We therefore build MG5 with smaller $dv/dz$ and $h$. From MG5, we further reduce $h$, slightly increase $dv/dz$ and build MG6. MG7 is similar to MG5, except that we set a higher INCL in order to explore the influence of INCL.

H$\alpha$ emission can suffer from dust attenuation. While dust can be present in the inner halo (which itself supports an internal origin for the EPG, \citealt{Howk00}), we expect that the most severe attenuation is produced by dust in the midplane, which would cause a drop in the H$\alpha$ intensity produced by portion of EPG located “behind” the disc with respect to the observer. To explore this effect, we use a extinction fraction $f$ to quantify the percentage of flux we observe from EPG located in the background of the midplane.  Based on the above simplified scenario, we build four more mock galaxies MG8--MG11, where $f=0.7$, 0.5, 0.25, 0.0, respectively; i.e., in MG8 we can still observe 70\% of the flux from EPG behind the disc while in MG11 we cannot see any emission from this EPG.

Table \ref{tab:mock} summarises the parameters we choose for the mock galaxies.

\begin{table*}
\resizebox{\textwidth}{!}{
\begin{tabular}{lrrrrrrrr}
\hline
\multicolumn{1}{c}{Mock galaxy} & \multicolumn{1}{c}{$h$} & \multicolumn{1}{c}{$dv/dz$} & \multicolumn{1}{c}{$v_z$} & \multicolumn{1}{c}{$v_R$} & \multicolumn{1}{c}{$\sigma$} &\multicolumn{1}{c}{$f_\mathrm{flux}$}&\multicolumn{1}{c}{N${_\mathrm{ind}}$}&\multicolumn{1}{c}{[S/N]$_{\mathrm{MED}}$} \\
  &  \multicolumn{1}{c}{[kpc]} & \multicolumn{1}{c}{[$\mathrm{km\,s^{-1}\,kpc^{-1}}$]} & \multicolumn{1}{c}{[$\mathrm{km\,s^{-1}}$]} 
 &\multicolumn{1}{c}{[$\mathrm{km\,s^{-1}}$]}&\multicolumn{1}{c}{[$\mathrm{km\,s^{-1}}$]}& &  \\ 
 \hline
 \multicolumn{1}{c}{(1)} & \multicolumn{1}{c}{(2)} & \multicolumn{1}{c}{(3)} & \multicolumn{1}{c}{(4)} & \multicolumn{1}{c}{(5)} & \multicolumn{1}{c}{(6)} & \multicolumn{1}{c}{(7)} & \multicolumn{1}{c}{(8)}&\multicolumn{1}{c}{(9)}  \\ 
 \hline \hline
  MG1 &$1.5$& $-40$ & $25$ & $25$ & $60$&0.30&1946&3.43\\ 
  MG2 &$1.5$& $-25$ & $25$ & $25$ & $50$&0.30&986&2.91\\ 
  MG3 &$1.5$& $-25$ & $25$ & $25$ & $50$&0.30&663&2.61\\ 
  MG4 &$1.5$& $-25$ & $-25$ & $-25$ & $50$&0.15&1125&2.88 \\ 
  MG5 &$1.0$& $-15$ & $-25$ & $-25$ & $50$&0.20&995&2.72\\ 
  MG6 &$0.5$& $-20$ & $-25$ & $-25$ & $50$&0.20&969&2.82 \\ 
  MG7 &$1.0$& $-15$ & $-25$ & $-25$ & $50$&0.15&876&2.71 \\ 
 \hline
  MG8 &$1.5$& $-25$ & $-25$ & $-25$ & $50$&0.30&1441&3.19 \\ 
  MG9 &$1.5$& $-25$ & $-25$ & $-25$ & $50$&0.30&1360&3.39 \\ 
  MG10 &$1.5$& $-25$ & $-25$ & $-25$ & $50$&0.30&1346&3.46 \\ 
  MG11 &$1.5$& $-25$ & $-25$ & $-25$ & $50$&0.30&1236&3.38 \\ 
\hline
\end{tabular}
}
\caption{Parameters for building mock galaxies. Columns: (1) Mock galaxy ID. (2)--(6): EPG parameters, see Sec.~\ref{sec:epgmodel}. (7) EPG flux fraction. {(8) Number of independent EPG voxels modelled, computed as the number of voxels in the masked datacube with intensity above twice the rms-noise divided by the number of voxels per resolution element. (9) The median S/N of the voxels in the EPG cube with intensities above twice the assumed rms noise.} 
All models use NGC~3982 as a template for the PA, INCL, systematic velocity, and rotation curve, although we note that MG7 has a different INCL of 45$^{\circ}$. In MG8--MG11 only a fraction ($f=0.7$,0.5,0.25,0.0) of emission from EPG located ``behind" the disc is visible. For all mock galaxies, $R_\mathrm{d}=1.6\,\mathrm{kpc}$, $h_\mathrm{d}=0.2\,\mathrm{kpc}$, $\sigma_\mathrm{d}=27\,\mathrm{km\,s^{-1}}$, $R_g=0.8\,\mathrm{kpc}$, and $\gamma=2.34$ (see Sec.~\ref{sec:epgmodel} and Sec.~\ref{sec:buildmock}).}
\label{tab:mock}
\end{table*}
}
\subsection{Mock galaxy test results }
\label{sec:mockresults}
We now analyse the mock galaxy datacubes built using the parameters listed in Table~\ref{tab:mock} with the method described in Sec.~\ref{sec:method}. We summarise the fitting results in Table~\ref{tab:mockresult}.  

\begin{figure*}
\centering
\includegraphics[scale=0.35]{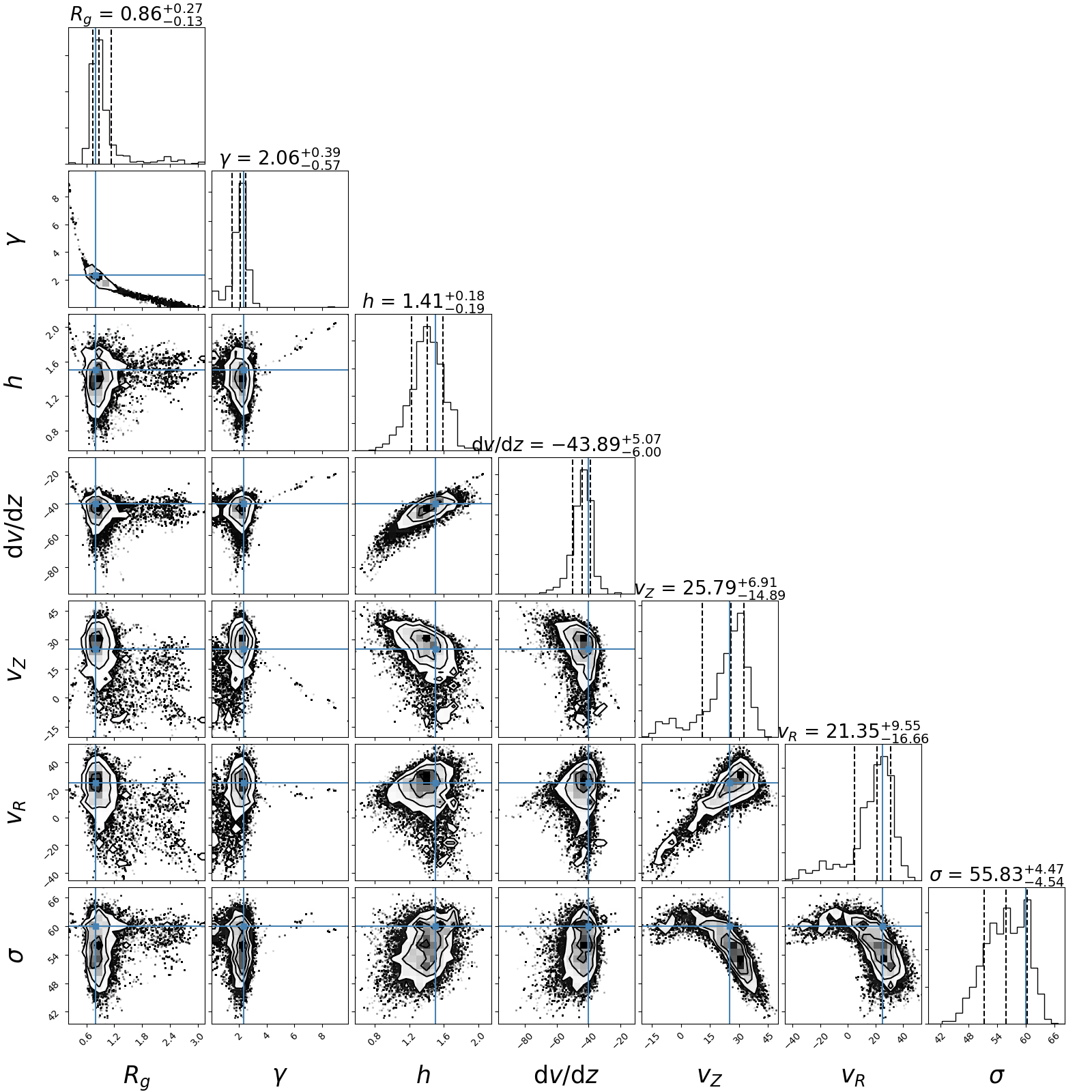}
\includegraphics[scale=0.35]{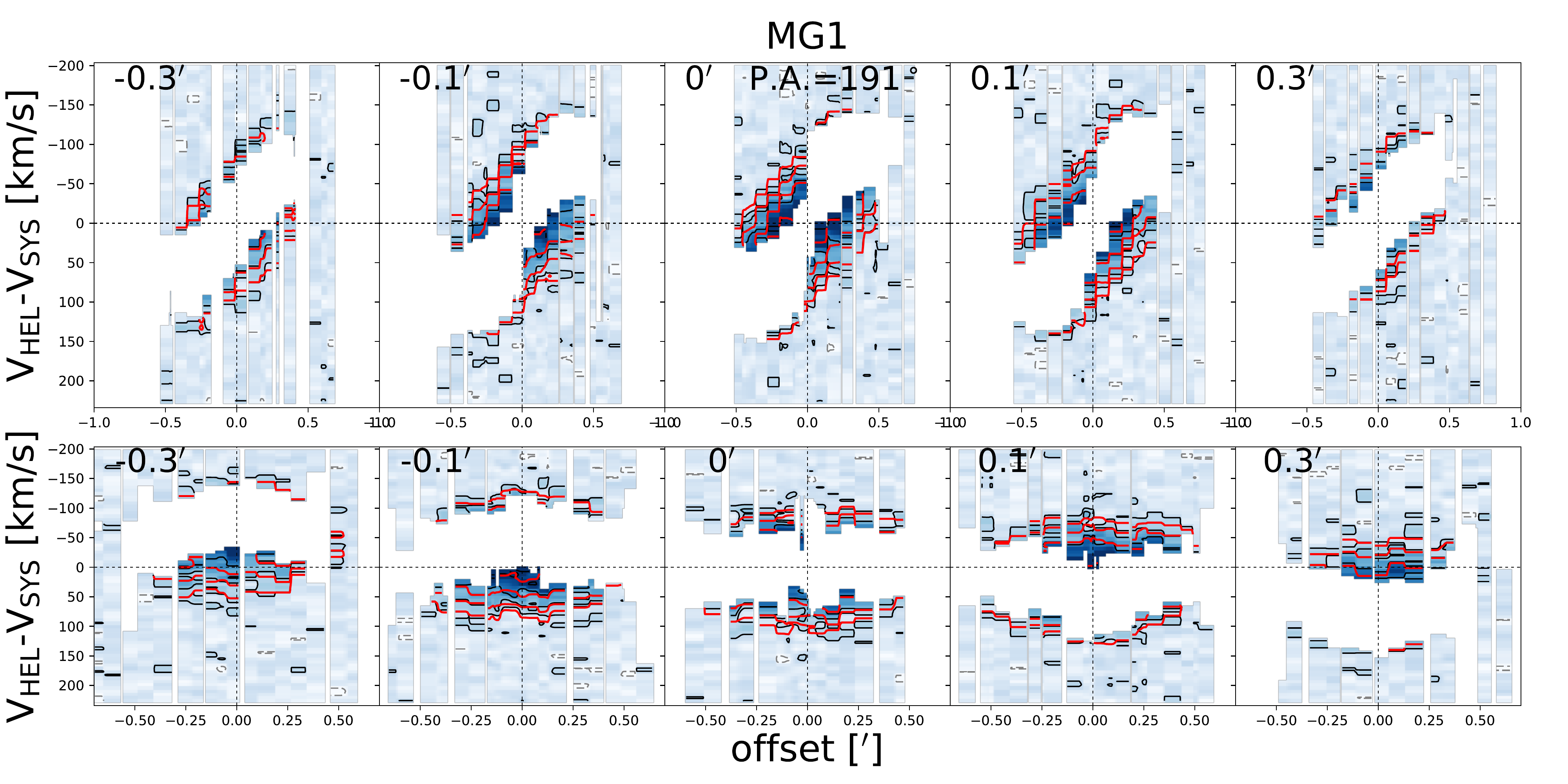}
\caption{MCMC fitting results for mock galaxy MG1. Top panel: corner plots showing the parameter correlations (density contours), their marginalised probability distributions (histograms on the top of density plots), and best-fit values. Best-fit values are defined as the median of the posterior distributions, and their uncertainties are taken as half the difference between the 84th and 16th percentiles of the distribution. Blue vertical and horizontal lines show the true values for these parameters. Bottom panel: position--velocity (pv) slices from the data (black contours) and from best-fit model (red contours). The white area represents internal mask. Top panels are pv slices along the major axis with offsets $-0\farcm3$, $-0\farcm1$, $0\farcm$, $0\farcm1$, $0\farcm3$. Bottom panels are pv slices along the minor axis with offsets $-0\farcm3$, $-0\farcm1$, $0\farcm$, $0\farcm1$, $0\farcm3$.}
\label{fig:corner1}
\end{figure*}

Figure~\ref{fig:corner1} shows the MCMC fitting result for MG1. For most parameters, the fit looks robust, except for the strong degeneracy between $R_g$ and $\gamma$, which also occurs in the \ion{H}{i} data analysis \citep{Marasco19}, and we will not comment on results of these two parameters for the rest of this section. We notice that the distribution of $v_R$ and $v_z$ have peaks consistent with input values but also show stronger tails towards one side that however do not influence the recovery of these parameters. The values of $h$, $dv/dz$, $v_z$, $v_R$, and $\sigma$ (marked with blue lines in Fig.~\ref{fig:corner1}) always fall within the 1$\sigma$ range  of the best-fit values, showing that the EPG parameters are recovered successfully.  The position--velocity (pv) slices of the best-fit model are also consistent with the mock data. We find a value of 0.32 for the flux fraction of EPG in MG1, which is derived as the total flux ratio between the EPG best-fit model and the datacube, overestimating the input value 0.3 by only 5\%. In general, we believe the code can reliably recover the EPG parameters in a case where the EPG emission is strong.

 The results for other mock galaxies are shown in Appendix~A. In MG2--MG7 the code works well for five EPG parameters ($h$, $dv/dz$, $v_z$, $v_R$, and $\sigma$), whose input values always fall within (or very close to) 1$\sigma$ range of the best-fit values, showing the reliability of our code for datacubes with even very low S/N or low EPG fractions. We also notice that the code tends to overestimate the flux fraction of EPG, which is not surprising given the poor estimation we have for $R_g$ and $\gamma$. 
 
 For MG8--MG10, the five EPG parameters also fall within (or close to) 1$\sigma$ range, except for $dv/dv$ in MG10, which is more than 2$\sigma$ away from the input value. The parameters (especially $dv/dz$,$v_z$, and $v_R$) are not well recovered in MG11, which is the case where we assume no EPG behind the disc is visible. This is an extreme assumption and is not likely to be the case for our galaxies (see the discussion in Sec~\ref{sec:limitations}). The above results suggest that dust extinction will not influence our analysis on EPG significantly, even in an extreme case where we can observe only 25\% emission of the EPG behind the disc.

\begin{table*}
\resizebox{\textwidth}{!}{
\begin{tabular}{lrrrrrrrr}
\hline
\multicolumn{1}{c}{Mock galaxy} &\multicolumn{1}{c}{$\Delta R_g$}&\multicolumn{1}{c}{$\Delta \gamma$}   &\multicolumn{1}{c}{$\Delta h$} & \multicolumn{1}{c}{$\Delta dv/dz$} & \multicolumn{1}{c}{$\Delta v_z$} & \multicolumn{1}{c}{$\Delta v_R$} & \multicolumn{1}{c}{$\Delta \sigma$}&\multicolumn{1}{c}{\(\frac{\Delta f_\mathrm{flux}}{f_\mathrm{flux}}\)} \\
  &\multicolumn{1}{c}{[kpc]}& & \multicolumn{1}{c}{[kpc]} & \multicolumn{1}{c}{[$\mathrm{km\,s^{-1}\,kpc^{-1}}$]} & \multicolumn{1}{c}{[$\mathrm{km\,s^{-1}}$]} 
 &\multicolumn{1}{c}{[$\mathrm{km\,s^{-1}}$]}&\multicolumn{1}{c}{[$\mathrm{km\,s^{-1}}$]} &\\ 
 \hline
 \multicolumn{1}{c}{(1)} & \multicolumn{1}{c}{(2)} & \multicolumn{1}{c}{(3)} & \multicolumn{1}{c}{(4)} & \multicolumn{1}{c}{(5)} & \multicolumn{1}{c}{(6)} & \multicolumn{1}{c}{(7)} & \multicolumn{1}{c}{(8)}& \multicolumn{1}{c}{(9)}  \\ 
 \hline \hline
  MG1 &0.5$\sigma$&$-$0.7$\sigma$ & $-$0.5$\sigma$ &$-$0.8$\sigma$ & 0.0$\sigma$&$-$0.4$\sigma$ &$-$1.0$\sigma$&5\%\\ 
  MG2&1.0$\sigma$&$-$1.2$\sigma$&0.0$\sigma$&$-$0.6$\sigma$&$-$0.5$\sigma$&$-$0.2$\sigma$&$-$0.9$\sigma$&13\%\\ 
  MG3&1.5$\sigma$&$-$1.9$\sigma$&$-$0.7$\sigma$&$-$1.2$\sigma$&$-$0.3$\sigma$&0.1$\sigma$&$-$1.9$\sigma$&26\% \\
  MG4&1.3$\sigma$&$-$1.4$\sigma$&1.0$\sigma$&1.2$\sigma$&1.2$\sigma$&0.5$\sigma$&$-$0.7$\sigma$&33\%\\ 
  MG5 &1.4$\sigma$&$-$1.8$\sigma$&0.4$\sigma$&$-$0.5$\sigma$&1.0$\sigma$&0.5$\sigma$&$-$0.9$\sigma$&30\% \\ 
  MG6 & 0.9$\sigma$&$-$2.4$\sigma$&0.3$\sigma$&$-$0.1$\sigma$&0.2$\sigma$&0.4$\sigma$&$-$1.6$\sigma$&30\% \\ 
  MG7 &1.0$\sigma$&$-$1.0$\sigma$&$-$0.6$\sigma$&$-$0.5$\sigma$&1.0$\sigma$&0.9$\sigma$&0.2$\sigma$&20\%\\ 
 \hline
  MG8 &1.1$\sigma$&$-$1.3$\sigma$&0.5$\sigma$&$-$0.4$\sigma$&0.9$\sigma$&0.3$\sigma$&$-$0.3$\sigma$&7\%\\ 
  MG9 &1.1$\sigma$&$-$1.4$\sigma$&$-$0.8$\sigma$&$-$1.5$\sigma$&$-$0.1$\sigma$&0.6$\sigma$&$-$0.7$\sigma$&3\%\\ 
  MG10 &0.6$\sigma$&$-$0.8$\sigma$&$-$1.3$\sigma$&$-$2.2$\sigma$&0.5$\sigma$&0.3$\sigma$&0.0$\sigma$&$-$3\%\\ 
  MG11 &3.6$\sigma$&$-$1.7$\sigma$&$-$1.9$\sigma$&$-$2.3$\sigma$&1.3$\sigma$&0.7$\sigma$&1.1$\sigma$&$-$20\%\\

\hline
\end{tabular}
}
\caption{Mock galaxy test results. Columns: (1) Mock galaxy ID. (2)--(8): The deviation of EPG parameters' best-fit results from input values. (9) The relative difference between best-fit models' flux fraction and input flux fraction.}
\label{tab:mockresult}
\end{table*}

Our mock galaxy tests are limited to this relatively small sample because the fitting is time consuming. The parameter space we explore may not cover the entire range of EPG properties in real galaxies. Nevertheless, the tests give us confidence in the following statements. (1) Implementing the EPG modelling code on H$\alpha$ data is possible, our modifications are reliable, and even moderate dust extinction does not influence our results significantly. (2) The code is still reliable for very low S/N of EPG (e.g.\ MG3) or when EPG flux fraction is as low as 15\%. (3) The code works well for five EPG parameters ($h$, $dv/dz$, $v_z$, $v_R$, and $\sigma$), whose input values always fall within (or close to) 1$\sigma$ range of the best-fit values. (4) Because of the intrinsic degeneracy between $R_g$ and $\gamma$, the code cannot in general reproduce these two parameters correctly. As a consequence, flux fraction of the EPG is not fully recovered either. But this does not affect the kinematic analysis, which is our prime goal here.

\section[Modelling the extraplanar gas of the DiskMass Halpha sample]{Modelling the extraplanar gas of the DiskMass H$\alpha$ sample}
\label{sec:results}

We now apply the method introduced in Sec.~\ref{sec:method} to determine the properties of the EPG in NGC~3982 and NGC~4152. 

\subsection{NGC~3982}
\label{sec:ngc3982}

\begin{figure}
\centering
\includegraphics[scale=0.2]{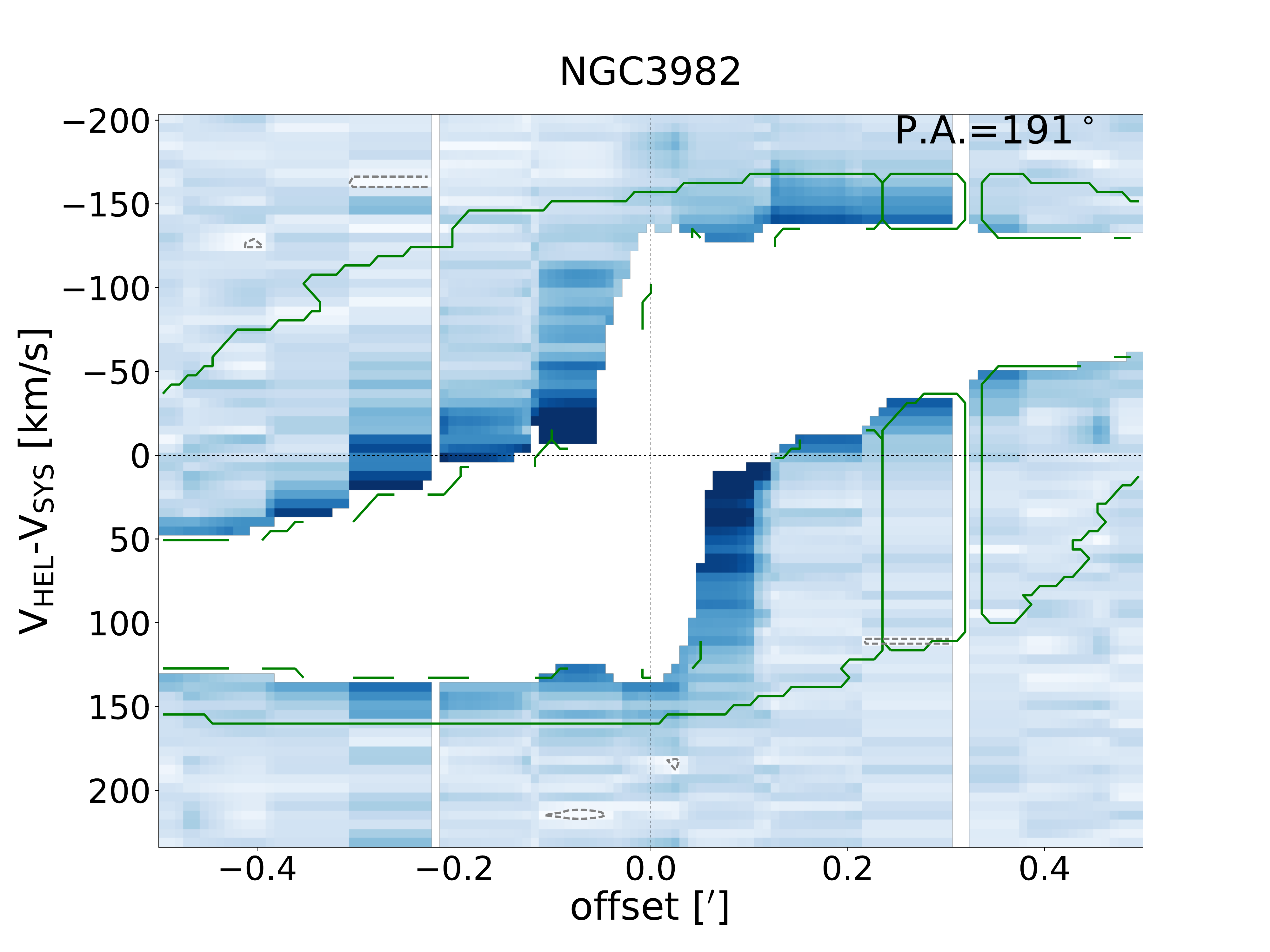}
\caption{Position--velocity slice along the major axis of NGC~3982 with external mask (green dashed contour) and internal mask (blank regions).}
\label{fig:maskplot}
\end{figure}

Following the method described in Sec.~\ref{sec:sep}, we extract EPG emission from the NGC~3982 data by implementing external and internal masks. These masks, as shown in Fig.~\ref{fig:maskplot}, seem to effectively separate EPG emission from the disc emission and pure noise. The internal mask shown in Fig.~\ref{fig:maskplot} is computed using an empirical mask threshold of $2\times\mathrm{rms}$ noise (see Sec.~\ref{sec:sep}). To explore the influence of the mask threshold on the fitting result, we also generate two additional internal masks using $1.5\times$ and $2.5\times\mathrm{rms}$ noise as thresholds.

\begin{figure*}
\centering
\includegraphics[scale=0.3]{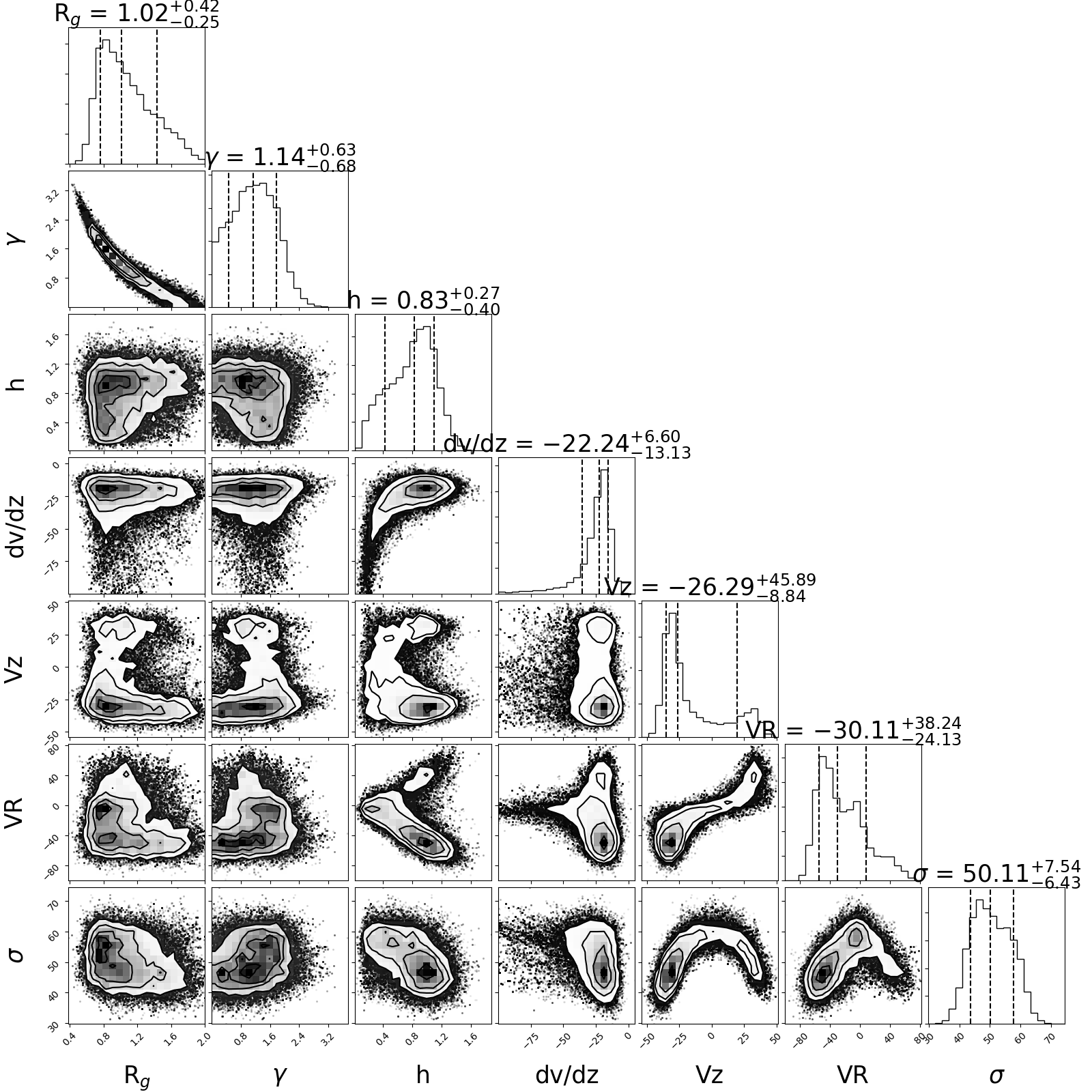}
\includegraphics[scale=0.5]{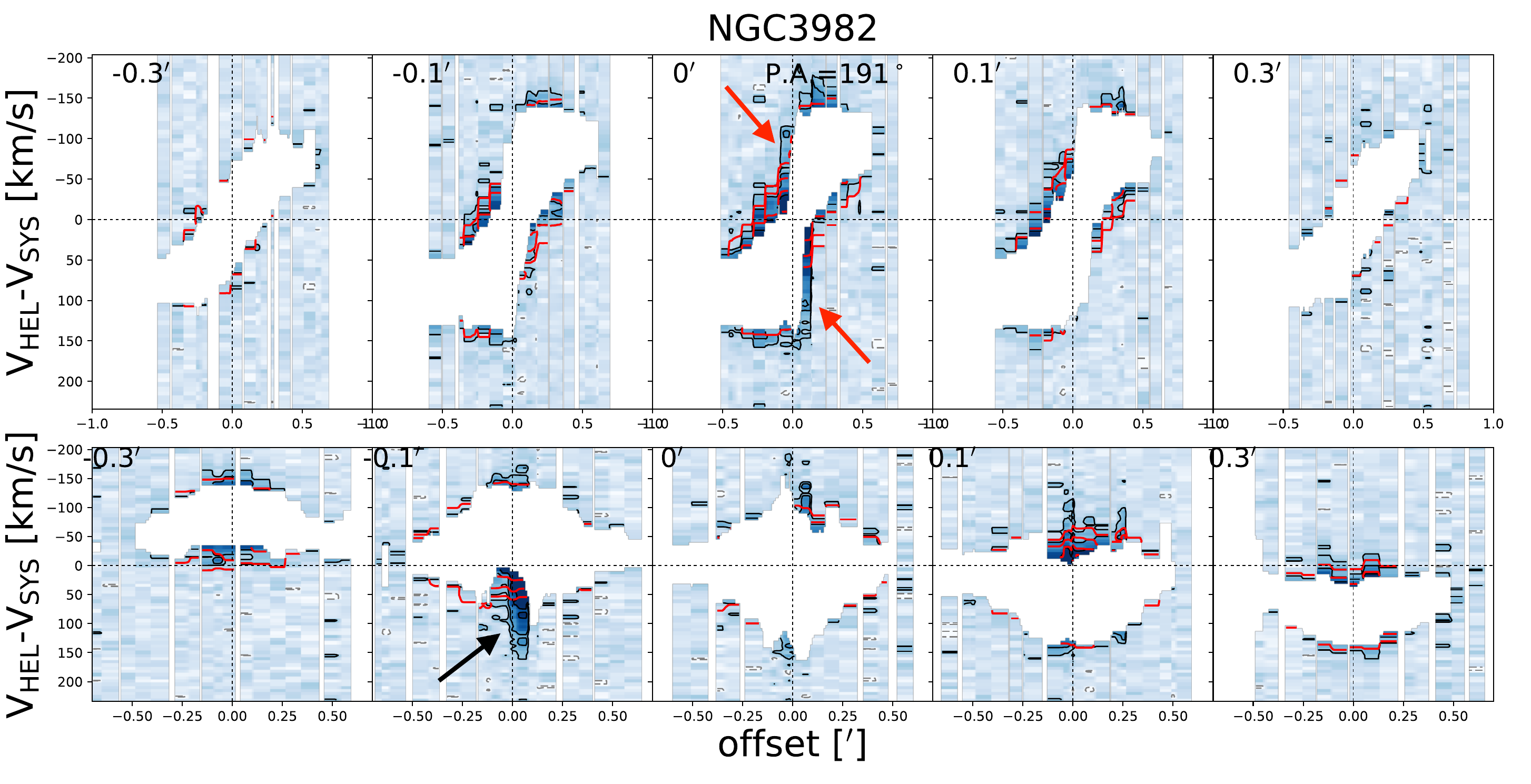}
\caption{Results for NGC~3982. Top panels: corner plots showing the parameter correlations (the density contours), their marginalised probability distributions (histograms at the top of density plots) and best-fit values from the MCMC fitting. Bottom panels: pv slices from the data (black contours) and the best-fit model (red contours). The white area represents the internal mask. Top panels are slices along major axis with offsets $-0\farcm3$, $-0\farcm1$, $0\farcm$, $0\farcm1$, $0\farcm3$. Bottom panels are slices along the minor axis with offsets $-0\farcm3$, $-0\farcm1$, $0\farcm$, $0\farcm1$, $0\farcm3$. A bright filament is indicated  by a black arrow, and the central region high-velocity parts are indicated by red arrows (see discussion in Sec~\ref{sec:ngc3982}).} 
\label{fig:ngc3982}
\end{figure*}

Once the rotation curve (derived from $\mathrm{^{3D}}$Barolo, see Sec.~\ref{sec:epgmodel} and Fig.~\ref{fig:rot}) and masks are ready, we fit the residual EPG signal. We show our results in Fig.~\ref{fig:ngc3982}. The pv slices from the data and the best-fit model are very consistent. There are some local features, however, which cannot be recovered by our smooth and axisymmetric EPG model intrinsically, for example the bright filament in the pv-slice $-0\farcm1$ offset from the minor axis (indicated by a black arrow). We also note the high-velocity part in the central region (pointed out by red arrows), which is possibly due to outflows from the AGN or non-circular motions caused by the nuclear bar (see \citealt{Knapen02}). Modelling such features is beyond the scope of this paper, although we discuss their influence on our fitting results in Sec.~\ref{sec:limitations}. The corner plots look robust, except for the expected degeneracy between $R_g$ and $\gamma$. The best-fit parameters are shown in Table~\ref{tab:result}. Our results indicate that there is abundant ionised EPG in NGC~3982, with lagging rotation, vertical and radial inflow, and large velocity dispersion\footnote{We note that the flux-weighted velocity dispersion should be lowered by a factor 0.7 with respect to the mass-weighted values quoted in Table~\ref{tab:result}, due to the difference between \ion{H}{i} and H$\alpha$ line profiles discussed in Sec.~\ref{sec:modifications}.}. We note the large uncertainties on the radial and vertical velocities and will discuss these below in Sec~\ref{sec:reliability}. Since our H$\alpha$ data is not flux calibrated (see above Sec~\ref{sec:diskmass}), the absolute flux of EPG is not available; consequently we can not conduct a flux-to-mass conversion and estimate the total mass of the EPG in NGC~3982.

For the other two internal masks, with thresholds $1.5\times\mathrm{rms}$ and $2.5\times\mathrm{rms}$,  we also implement the MCMC fitting method. We find that using these thresholds leads to best-fit parameters which are perfectly compatible with those reported in Table~\ref{tab:result} (Sec.~\ref{sec:limitations}).

\begin{table*}
\resizebox{\textwidth}{!}
{
\begin{tabular}{lrrrrrrrrrr}
\hline
\multicolumn{1}{c}{Galaxy name}  & \multicolumn{1}{c}{$R_g$}& \multicolumn{1}{c}{$\gamma$}& \multicolumn{1}{c}{$h$} & \multicolumn{1}{c}{$dv/dz$} & \multicolumn{1}{c}{$v_z$} & \multicolumn{1}{c}{$v_R$} & \multicolumn{1}{c}{$\sigma$} &\multicolumn{1}{c}{$f_{\mathrm{flux}}$}&\multicolumn{1}{c}{N$_\mathrm{ind}$}&\multicolumn{1}{c}{[S/N]$_\mathrm{MED}$} \\
  &  \multicolumn{1}{c}{[kpc]} &  \multicolumn{1}{c}{[kpc]} & \multicolumn{1}{c}{[$\mathrm{km\,s^{-1}\,kpc^{-1}}$]} & \multicolumn{1}{c}{[$\mathrm{km\,s^{-1}}$]} 
 &\multicolumn{1}{c}{[$\mathrm{km\,s^{-1}}$]}&\multicolumn{1}{c}{[$\mathrm{km\,s^{-1}}$]}& & & &\\ 
 \hline
 \multicolumn{1}{c}{(1)} & \multicolumn{1}{c}{(2)} & \multicolumn{1}{c}{(3)} & \multicolumn{1}{c}{(4)} & \multicolumn{1}{c}{(5)} & \multicolumn{1}{c}{(6)} & \multicolumn{1}{c}{(7)} & \multicolumn{1}{c}{(8)} & \multicolumn{1}{c}{(9)}& \multicolumn{1}{c}{(10)}& \multicolumn{1}{c}{(11)}\\ 
 \hline \hline
  NGC~3982 &1.02$^{+0.42}_{-0.25}$&1.14$^{+0.63}_{-0.68}$&0.83$^{+0.27}_{-0.40}$&$-$22.24$^{+6.60}_{-13.13}$ &$-$26.69$^{+45.89}_{-8.84}$&$-$30.11$^{+38.24}_{-24.13}$ & 50.11$^{+7.54}_{-6.43}$&27\% &929&2.88\\ 
\\
  NGC~4152&3.61$^{+1.79}_{-1.05}$&0.38$^{+0.49}_{-0.27}$ &1.87$^{+0.43}_{-0.56}$&$-$11.18$^{+3.49\phantom{0}}_{-4.06\phantom{0}}$ &$-$25.14$^{+35.36}_{-11.67}$ &$-$27.96$^{+50.86}_{-25.95}$ & 59.10$^{+8.09}_{-8.28}$&15\% &714&2.83\\
\hline
\end{tabular}}

\caption{MCMC fitting results for NGC~3982 and NGC~4152.  Columns: (1) Galaxy name. (2)--(8): EPG parameters. (9) EPG flux fraction. (10) Number of independent EPG voxels modelled, computed as the number of voxels in the masked datacube with intensity above twice the rms-noise divided by the number of voxels per resolution element. (11) The median S/N of the voxels in the EPG cube with intensities above twice the assumed rms noise.}
\label{tab:result}
\end{table*}

\subsection{NGC~4152}
\label{sec:ugc7169}
We apply the method discussed in Sec.~\ref{sec:ngc3982} to the NGC~4152 H$\alpha$ datacube. The fitting results for NGC~4152 are shown in Fig.~\ref{fig:ugc7169}. We note that, similarly to NGC~3982, there are also central regions (indicated by red arrows) with high-velocity that cannot be reproduced by the EPG model. Another irregular pattern is shown in the pv-slice offset $-0\farcm1$ from the major axis, which seems to be a local feature than cannot be reproduced by our EPG model intrinsically (indicated by a black arrow). Best-fit values are shown in Table~\ref{tab:result}. The best-fit values for NGC~4152  suggest that this galaxy also has ionised EPG with lagging rotation, radial and vertical inflow, and large velocity dispersion.

\begin{figure*}
\centering
\includegraphics[scale=0.3]{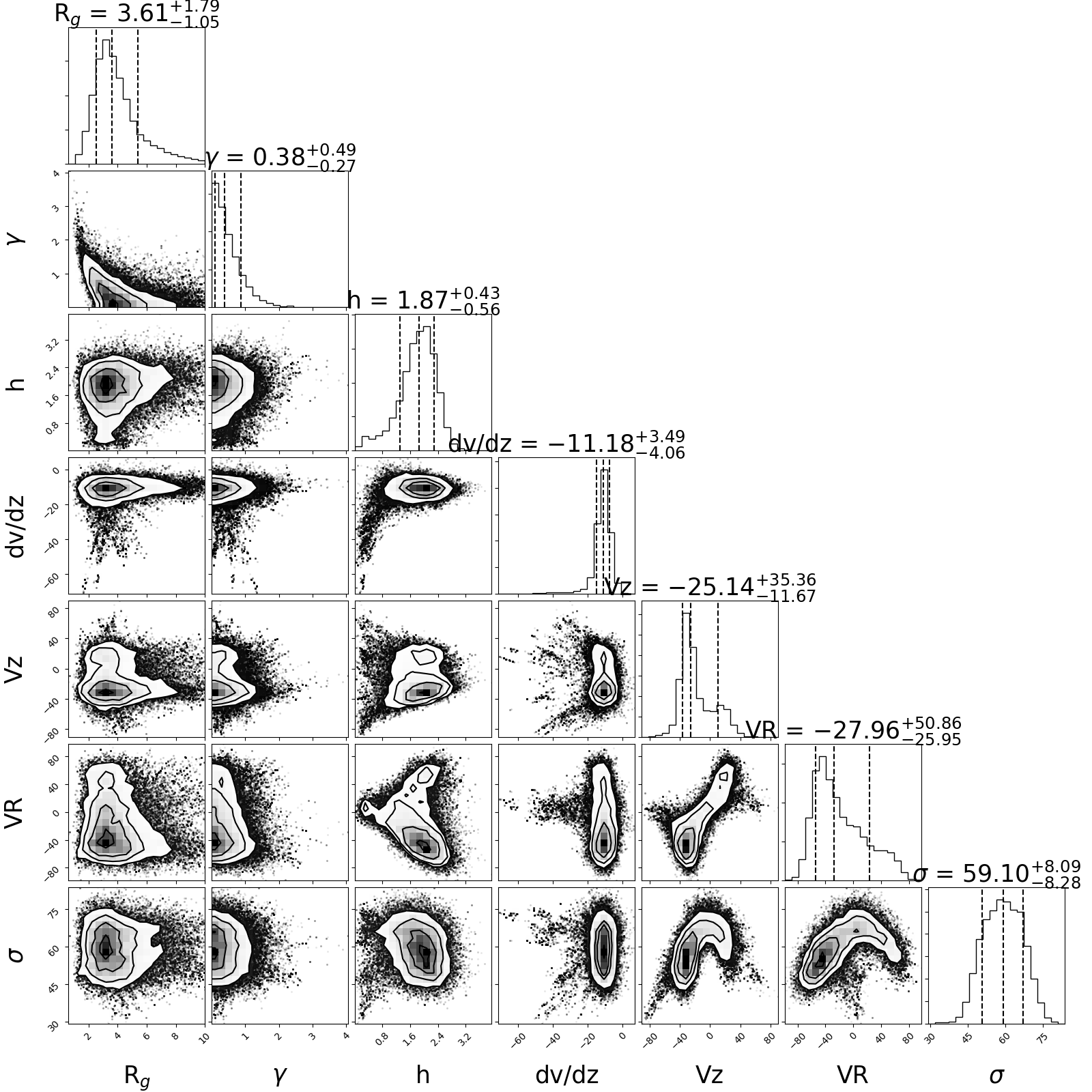}
\includegraphics[scale=0.5]{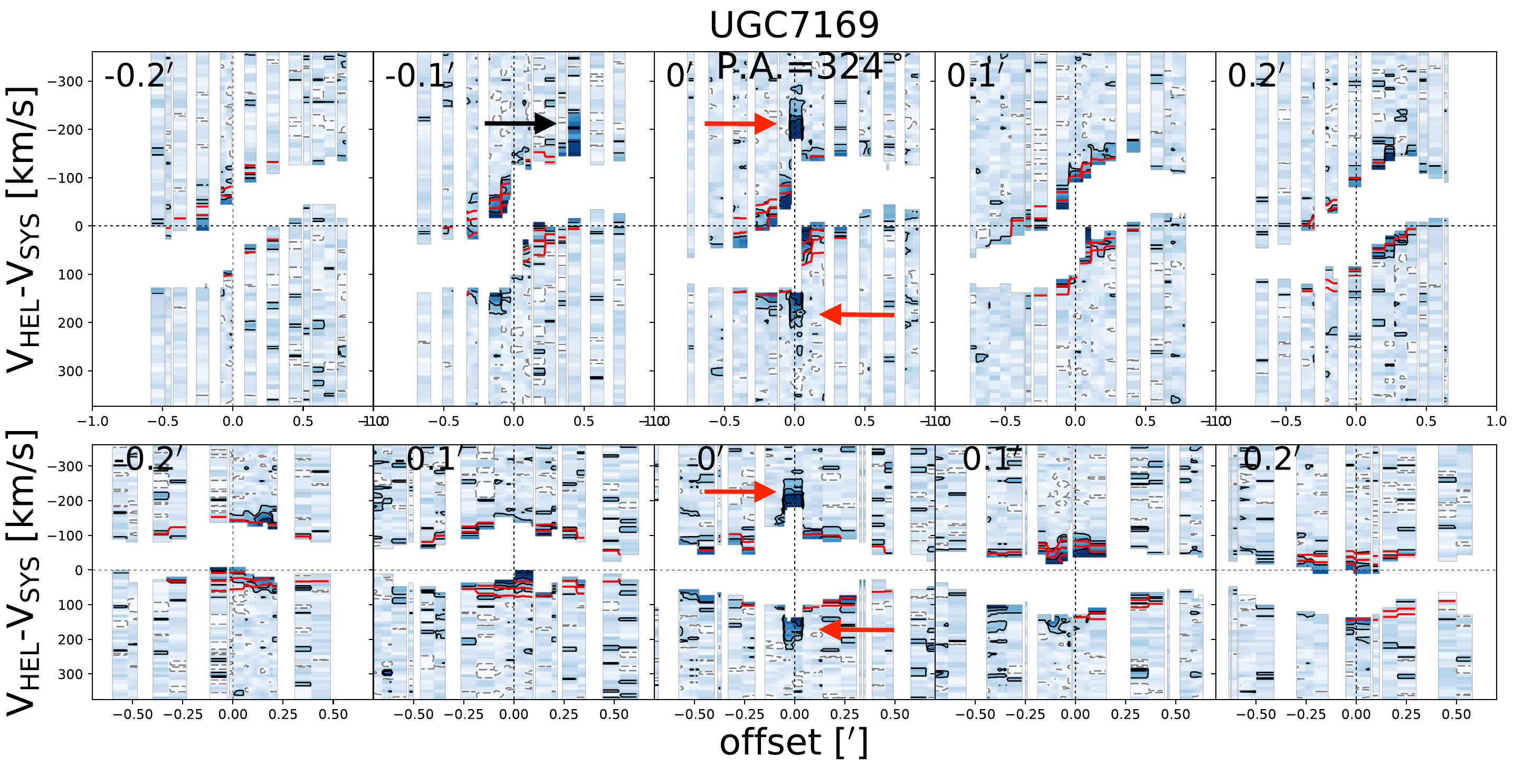}
\caption{As in Fig.~\ref{fig:ngc3982}, but for NGC 4152.}
\label{fig:ugc7169}
\end{figure*}

\section{Discussion}
\label{sec:discussion}
\subsection{Limitations of our EPG model}
\label{sec:limitations}
The EPG modelling code introduced in \citet{Marasco19} allows us to study ionised EPG with 3D modelling for the first time. Limitations of the code are discussed in detail in Sec.~4.1 of \citet{Marasco19}. Below we discuss some limitations and their influences on our analysis of ionised EPG.

The code assumes a smooth and symmetric (both axisymmetric and across the galactic plane) EPG model which is not the case for all galaxies. Indeed, both NGC~3982 and NGC~4152 show local features that cannot be fitted by our EPG model (see Sec.~\ref{sec:results}). We note a slight trend that the model tends to ‘overshoot’ the normal EPG emission component, indicating that the model attempts to recover local features. Nevertheless, those local features do not influence the fitting results significantly. Apart from those local features, our data seem to be symmetric and smooth. We have also attempted to exclude some of the local features by hand (i.e., a manual mask) and the fitting results are compatible.

Another assumption made by the model is constant PA and INCL for the galaxy. Galaxies are typically regular within their effective radius (e.g.\ approximately within the size of their star-forming disc), while warps and other asymmetries are more significant at larger radii. In the DMS sample, only galaxies with regular morphology and kinematics of H$\alpha$ data are selected \citep{Bershady10}. We therefore conclude that constant PA and INCL is a well motivated assumption.

Ionised gas suffers from dust extinction, of course \citep{2006agna.book.....O}, which, however, is not included in our model. Nevertheless, mock galaxy tests (MG8--MG10) show that our code can recover EPG parameters well for mock datacubes where ionised EPG located behind the disc is only partly (down to $f=25\%$) visible due to extinction. But in an extreme case (MG11), where gas behind the disc is completely invisible,  EPG parameters are \emph{not} well recovered. Luminous star-forming galaxies like NGC~3982 and NGC~4152 typically have extinction $A_V < 1.0$ \citep{Calzetti01} in most regions, which corresponding to $f \sim 40\%$. We therefore conclude that our results are reliable for these two galaxies. When spatially resolved dust extinction is available (e.g.\ from H$\beta$/H$\alpha$ line ratios) in future studies, they could be included in our model.

An important step in our method is to separate the EPG and the disc gas, for which we use an empirical threshold ($2\times\,\mathrm{rms\ noise}$) to mask out disc emission. In principle mask threshold could influence the EPG residuals and thus the fitting results. Our experiment in Sec.~\ref{sec:ngc3982} shows that the fitting results is not sensitive to the mask threshold. 

\subsection{Reliability of the fitting results}
\label{sec:reliability}
To characterise the EPG in our galaxies we fit a model with seven independent parameters to the data. Given the number of parameters and the quality of our data, one may question the reliability of our results. The mock galaxy tests in Sec.~\ref{sec:mock} provide some insights into this question. Comparing the S/N, EPG parameters and flux fractions of NGC~3982 and NGC~4152 with mock galaxies, we find that they have comparable S/N, flux fraction, and EPG parameters. We conclude that these mock galaxies MG1-MG10 can well represent NGC~3982 and NGC~4152. We do not include MG11, since it assumes an extreme case of (complete) dust extinction and is therefore not comparable with our sample.

{From the mock galaxy tests we find that our method does not recover $R_g$ and $\gamma$ well} due to the intrinsic degeneracy between these two parameters. We therefore should not trust the $R_g$ and $\gamma$ values inferred for NGC~3982 and NGC~4152. In the following we will therefore not discuss these two parameters.

The reliability of the estimated EPG scale height $h$ is influenced by the inclination (INCL) of the galaxy. For example, it is easier to measure the scale height of EPG in an edge-on galaxy than in a galaxy with intermediate inclination. The estimated INCL of NGC~3982 is 26\fdg2, which raises the question if our estimate for scale height $h$ is still accurate at such low inclination. With the same INCL (except MG7) as NGC~3982, the scale heights $h$ of MG1--MG10 are recovered correctly (within 1$\sigma$ range of best-fit value in MG1--MG9, and only 1.3$\sigma$ away in MG10), suggesting that our scale height $h$ values for NGC~3982 (even with this low inclination angle) and NGC~4152 are reliable.

Our method also recovers the velocity gradient $dv/dz$ well. In most cases the differences between best-fit values and input values of $dv/dz$ are less than (or close to) $1\sigma$. We note that the code overestimates the velocity gradient by more than 1.5$\sigma$ in MG9 and MG10, where we introduce dust extinction. We conclude that NGC~3982 and NGC~4152 undoubtedly show a lagging rotation, but perhaps with a smaller velocity gradient than suggested by the best-fit values if dust extinction in their discs are significant.

As for the radial velocity $v_R$ and vertical velocity $v_z$, our method manages to recover them correctly (within or very close to $1\sigma$). Although these are successfully recovered in mock galaxy tests, we note that the error bars for $v_R$ and $v_z$ in NGC~3982 and NGC~4152 are relatively large, even drifting towards zero velocity. From the corner plots in Figs.~\ref{fig:ngc3982} and \ref{fig:ugc7169}, we find that the peaks of $v_R$ and $v_z$ are clearly located at the negative value side, and it only is the small tail to the positive side that results in the large error-bar. Such tails also appear in mock galaxy tests (see, e.g., Fig.~\ref{fig:corner1}). We conclude that the code indeed prefers radial and vertical inflows for NGC~3982 and NGC~4152, but with these large error-bars, we also cannot completely exclude the possibility of zero radial and/or vertical EPG velocities.

\subsection{Comparison with previous work}
\label{sec:previouswork}
We now compare our ionised EPG parameters to those previously
measured in the literature. 

\citet{Rossa03a} detected ionised EPG in 30 edge-on galaxies with photometric observations and found a typical scale height $h$ of 1--$2\,\mathrm{kpc}$. Another photometric study conducted by \citet{Miller03a} reported scale heights of ionised EPG varying from $0.4\,\mathrm{kpc}$ to $17.9\,\mathrm{kpc}$ for 16 edge-on galaxies, with a mean value of $4.3\,\mathrm{kpc}$. This relatively high value of scale height is likely due to the deep flux limit in their data. Using a MaNGA sample of 65 edge-on galaxies, \citet{Bizyaev17} derived a median scale height of $1.2\pm0.5\,\mathrm{kpc}$ for ionised EPG. The EDGE-CALIFA Survey used a sample of 25 edge-on galaxies and measured a median scale height of $0.8^{+0.7}_{-0.4}\,\mathrm{kpc}$ \citep{Levy19}. They also summarised a list of previous studies on ionised EPG and found a median scale height of $1.0\pm2.2\,\mathrm{kpc}$. The scale heights we derive for NGC~3982 ($0.83^{+0.27}_{-0.40}\,\mathrm{kpc}$) and NGC~4152 ($1.87^{+0.43}_{-0.56}\,\mathrm{kpc}$) are consistent with these studies. 

The velocity gradients of ionised EPG was also studied in several spectroscopic studies. For example, \citet{Heald05} reported a velocity gradient of $-15\,\mathrm{km\,s^{-1}\,kpc^{-1}}$ for ionised EPG in NGC~891 and $-8\,\mathrm{km\,s^{-1}\,kpc^{-1}}$ for NGC~5775. And the EDGE-CALIFA survey found a velocity gradient of $-21\,\mathrm{km\,s^{-1}\,kpc^{-1}}$ \citep{Levy19}.  The velocity gradients of NGC~3982 ($dv/dz=-22.24^{+6.60}_{-13.13}\,\mathrm{km\,s^{-1}\,kpc^{-1}}$) and NGC~4152 ($dv/dz=-11.18^{+3.49}_{-4.06}\,\mathrm{km\,s^{-1}\,kpc^{-1}}$) are also consistent with previous studies. 

Those previous measurements of EPG properties are from edge-on galaxies where the scale height can be directly measured and velocity gradients are calculated by analysing velocity profiles. It is noteworthy that our results for scale height and velocity gradient are very consistent with these studies although we use a completely different method, which is 3D modelling of the datacube. This method allows us to study a much broader sample of galaxies, as geometrical constraints become significantly less important.

Our analysis for NGC~3982 and NGC~4152 suggests that ionised EPG shows radial and vertical inflow, which has also been found in previous studies on other galaxies. The ionised EPG in NGC~2403 shows evidence of localised outflows, but the inflow seems to be diffused over the scales of the entire star-forming disc \citep{Fraternali04}. \citet{Zheng17} found  disc-wide ionised gas inflow in M33 via absorption line study against bright UV sources in the disc. Warm gas in the Milky Way traced by absorption lines has more detections at negative than positive velocities (\citealt{Lehner12}; see also left panel of Fig.4 in \citealt{Marasco13}), indicating that inflow is more prominent than outflow for ionised EPG in the Milky Way.  Other previous studies were limited to edge-on samples as mentioned above and lack simultaneous vertical and radial velocity analysis. But the inflow of \ion{H}{i} EPG has been found in many studies \citep[e.g.][]{Fraternali01,Marasco19}.

\subsection[Is the EPG produced by a galactic fountain?]{Is the EPG produced by a galactic fountain?}
In the classic galactic fountain scenario, the EPG originates from the disc, is pushed up by star-formation activity (stellar wind and supernova explosions), and eventually falls back to the disc (e.g.\ \citealt{Bregman80}). Therefore the gas is outflowing in the first stage and inflowing in the second stage. However, \ion{H}{i} EPG is usually found to be preferentially inflowing in observations. A plausible explanation is that the outflowing gas might be highly ionised and only cools down near the apocentre, thus \ion{H}{I} is only partly visible during the ascending phase and fully visible during the descending phase (e.g.\  \citealt{Fraternali06,Fraternali08,Marasco12}). In the above mentioned scenario, we would expect to see outflows to dominate the ionised gas kinematics, contrary to what we infer for the ionised EPG in NGC~3982 and NGC~4152.  

In what follows we discuss the possible explanations for this discrepancy, either within the galactic fountain framework or outside of it. One possible mechanism to explain this discrepancy is that the outflowing gas, being highly ionised, is at a temperature of around $10^5\,\mathrm{K}$ or more, in which case the gas is too hot to be seen in H$\alpha$ but may be visible through the absorption of \ion{O}{vi} or even X-ray emission (e.g.\ \citealt{Fraternali02b}). Later the hot gas cools down and turns into warm gas and neutral gas. Furthermore, it is likely that part of the cooling \ion{H}{i} gas is ionised by photons leaking from the disc again during the descending phase, and emits in H$\alpha$, which can explain why we see similar kinematic properties for the ionised (H$\alpha$) and neutral (\ion{H}{i}) EPG.

In Sec.~\ref{sec:introduction} we mention that a model of pure galactic fountain cannot fully explain the properties of EPG, in particular its slowly rotating and inflowing kinematics require fountain-driven condensation of the hot corona \citep{Fraternali17}. Hydrodynamic simulations are required to fully capture the physics of this mechanism \citep{Marinacci11,Armillotta16,Gronnow18}. When fountain gas moves through the hot corona, it mixes with the hot gas. The mixture increases the metallicity and decreases the temperature of the hot corona and therefore decreases its cooling time, making part of the hot corona condense into warm or cold gas. A sketch of this process can be seen in Fig.~1 of \citet{Fraternali13}. It is possible that this accretion mechanism can also explain the ionised EPG inflow found in our data, since the amount of warm EPG increases due to condensation in the descending part (also see Fig.~1 of \citealt{Fraternali13} and Fig.~1 of \citealt{Marasco13}).

While our model assigns a unique value to the inflow velocity of the EPG, the galactic fountain kinematics can be very complex. We stress that a preference for inflow does not mean that there is no outflow. It is likely that the inflow of ionised EPG due to condensation of the corona together with the \ion{H}{i} EPG being ionised in the descending phase leads to our models to prefer inflow for H$\alpha$ EPG. Moreover, as mentioned in Sec.~\ref{sec:previouswork}, outflowing gas is likely to be more localised (e.g.\ \citealt{Fraternali04, Chisholm16,Boettcher17}) since it comes from \ion{H}{ii} regions, which are clumpy and preferentially confined within the spiral arms. As fountain clouds pass through the halo they fragment and expand \citep{Marinacci11}, eventually producing a smooth and diffused rain of multi-phase gas onto the disc. In the above scenario, our model, which is based on the large-scale smooth emission, would also prefer the inflow. It is, however, not possible to test the above hypothesis with our current kinematic model. To better resolve this issue, a dynamical model for ionised EPG data is required and it will be subject of future investigations. 

Another possible explanation for the observed inflow is that of an external origin for the ionised EPG. In this scenario the H$\alpha$-emitting material would not originate from the circulation of fountain clouds, but it would rather trace gas accretion from the CGM \citep{Kaufmann06,Zheng17} produced either by the spontaneous cooling of the hot corona \citep[e.g.][]{Voit15,Sormani19} or by cold filaments penetrating the hot gas down to the disc \citep[e.g.][]{Fernandez12, Mandelker19}. However, we stress that the dominating motion in our EPG is not the inflow but the rotation, which has remarkable proximity to the disc with only a mild vertical gradient. This indicates that the accreting gas close to the disc must have a relatively high angular momentum \citep{Pezzulli16}, and/or be efficiently accelerated by the galactic fountain \citep{Marinacci11}.

\section{Conclusion}
 \label{sec:summarise}
In this paper we have studied the kinematic properties of ionised EPG in two galaxies from the DiskMass survey, NGC~3982 and NGC~4152. We have modelled the H$\alpha$ datacubes of these galaxies using the 3D kinematic modelling code from \citet{Marasco19}, which first separates the EPG emission from the disc emission and then fits the EPG residual with a model controlled by seven parameters via a Bayesian MCMC approach. The seven parameters characterising EPG properties are the scale length $R_g$, an exponential parameter $\gamma$ (see Eq.~\ref{eq:surface}), the scale height $h$, the velocity gradient $dv/dz$, the radial velocity $v_R$, the vertical velocity $v_z$, and the velocity dispersion $\sigma$. We have tested the reliability of our method with mock galaxies, which we have built assuming ionised gas properties compatible with those of  NGC~3982 and NGC~4152.

Our results for NGC~3982 and NGC~4152 can be summarised as follows. 
\begin{enumerate}
    \item  We have detected ionised EPG in both galaxies, with flux fraction between the EPG and the regularly rotating gas within the disc of 27\% (NGC~3982) and 15\% (NGC~4152).
    \item The scale heights for ionised EPG are $0.83^{+0.27}_{-0.40}\,\mathrm{kpc}$ (NGC~3982) and $1.87^{+0.43}_{-0.56}\,\mathrm{kpc}$ (NGC~4152), consistent with previous determination of ionised EPG in other systems. 
    \item The ionised EPG in both galaxies show lagging rotation,  with rotational gradients of $-22.24^{+6.60}_{-13.13} \,\mathrm{km\,s^{-1}\,kpc^{-1}}$ (NGC~3982) and $-11.18^{+3.49}_{-4.06}\,\mathrm{km\,s^{-1}\,kpc^{-1}}$ (NGC~4152).
    \item Our best-fit model suggests radial and vertical inflow for both galaxies. But we cannot exclude the possibility of zero radial and/or vertical velocity as the uncertainties for these two parameters are large.
    \item The velocity dispersions of ionised EPG are 50.11$^{+7.54}_{-6.43}\,\mathrm{km\,s^{-1}}$ (NGC~3982) and 59.10$^{+8.09}_{-8.28}\,\mathrm{km\,s^{-1}}$ (NGC~4152), roughly a factor of two larger than the disc values.
\end{enumerate}
The kinematic signature that we have derived is consistent with the EPG layer being mostly produced by galactic fountains, but gas accretion, either induced or spontaneous, may also play an important role. 
A dynamical model of EPG is needed to better understand the origin of the ionised EPG. A large sample of galaxies with better spatial resolution is also essential for further studies of ionised EPG. In the near future, WEAVE will provide IFU data for a large sample of late-type galaxies at a comparable spectral resolution, a wider wavelength range, and a higher angular resolution, which will enable a more comprehensive study of the ionised EPG of disc galaxies.

\section*{Acknowledgements}
We thank the referee for useful comments which helped improving the manuscript. The authors would like to thank the DiskMass Survey team for useful discussions and for making the SparsePak data available to us prior to public release. AL and MV would like to thank J.M. van der Hulst for useful suggestions.
AL was supported by Netherlands Research School for Astronomy (Nederlandse Onderzoekschool voor Astronomie, NOVA), Network 1, Project 10.1.5.9 WEAVE.
AM acknowledges the support by INAF/Frontiera through the "Progetti Premiali" funding scheme of the Italian Ministry of Education, University, and Research.
MV acknowledges financial support from the Netherlands Foundation for Scientific Research (NWO) through VICI grant 016.130.338.

\section*{Data Availability}

The data underlying this article were provided by the DiskMass Survey team by permission. Data will be made available upon reasonable request to the corresponding author with permission of the DiskMass Survey team.
\bibliographystyle{mnras}
\bibliography{references}
\appendix
\renewcommand\thefigure{\Alph{section}\arabic{figure}} 
\section{Mock Galaxy Test Results}
\label{appen:a}
Figure~\ref{fig:mg2} to Figure~\ref{fig:mg11} show the MCMC fitting results for MG2 to MG11. Discussion of these results can be found in Sec.~\ref{sec:mockresults}.
\begin{figure*}
\centering
\includegraphics[scale=0.3]{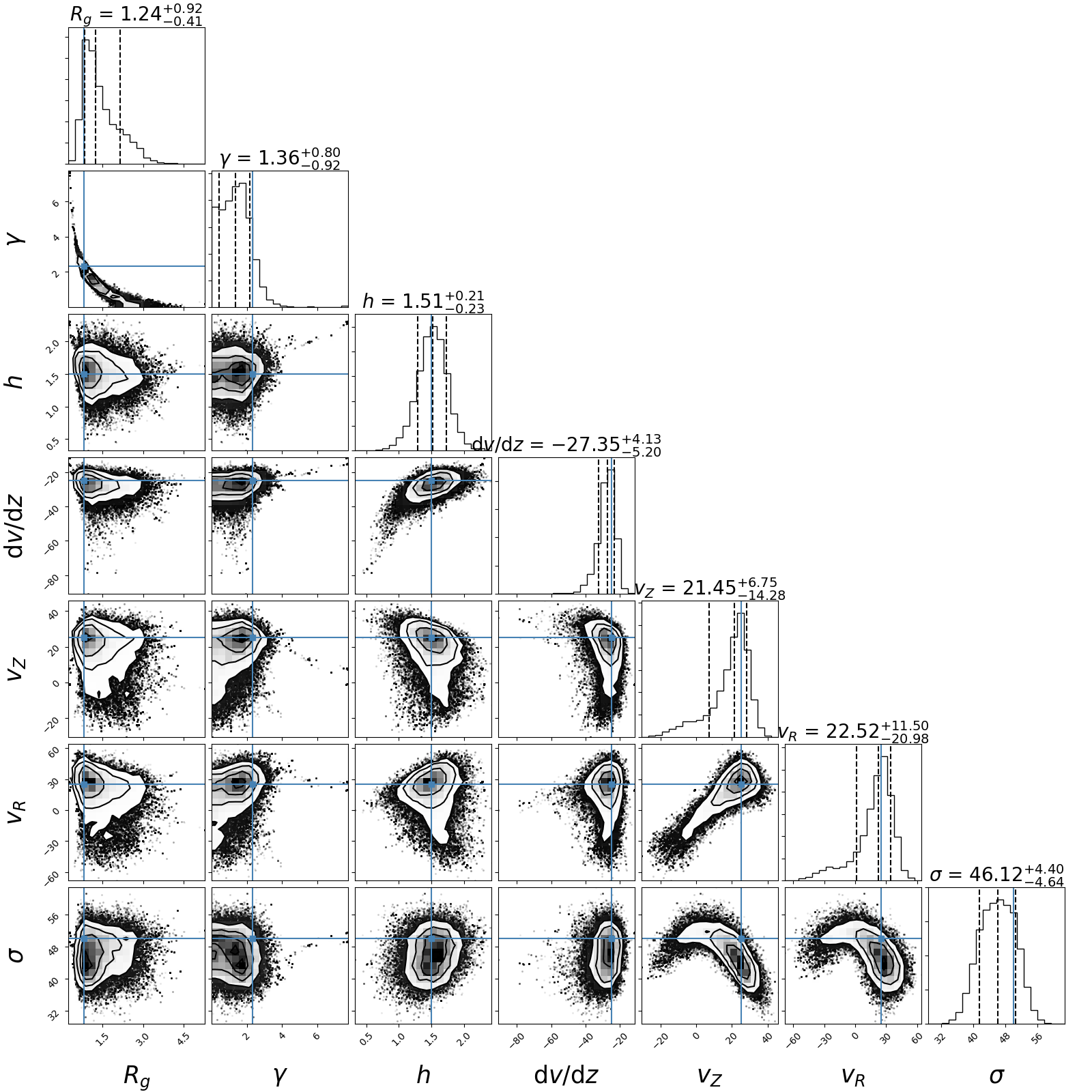}
\includegraphics[scale=0.3]{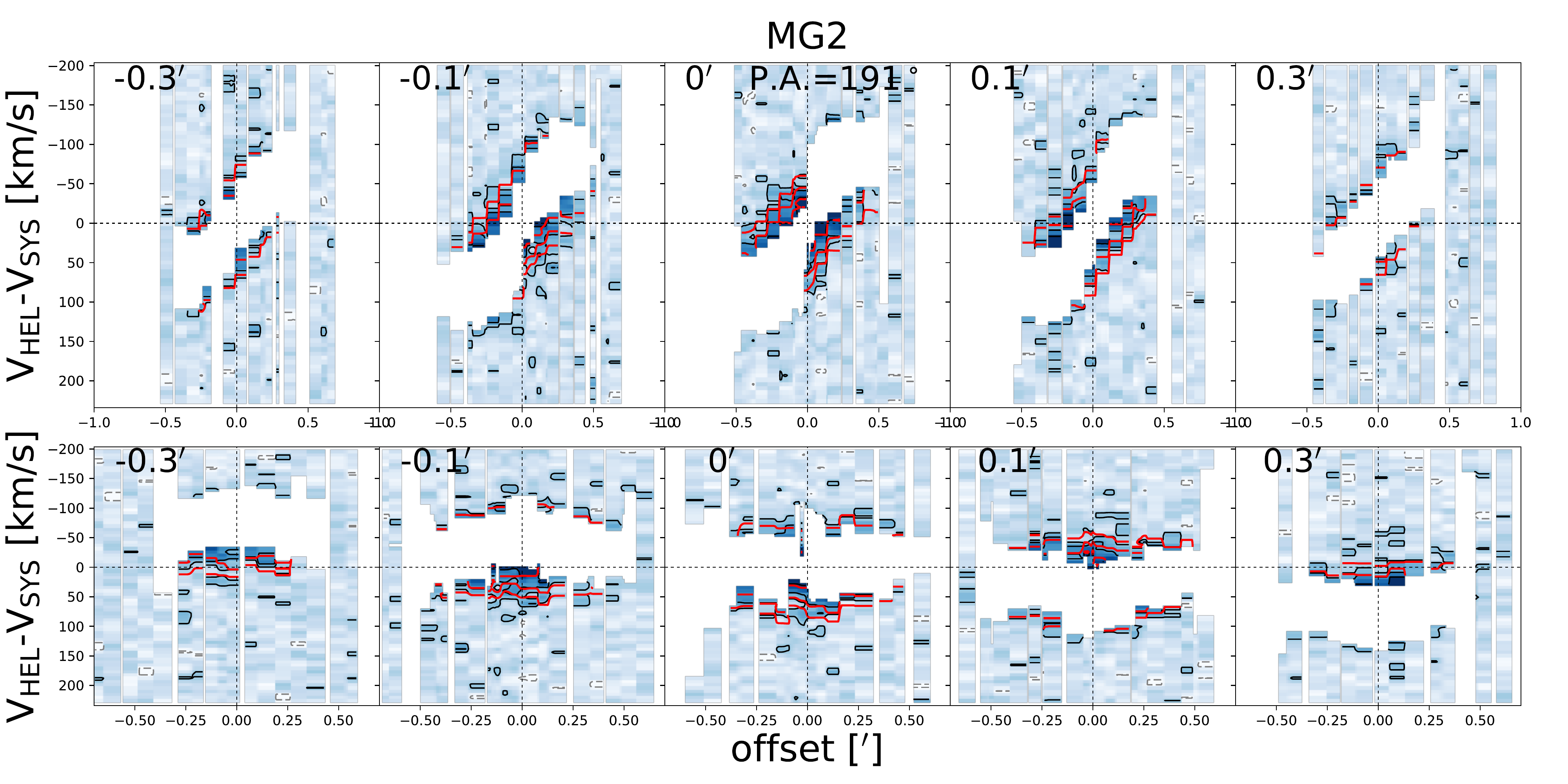}
\setcounter{figure}{0}
\caption{MCMC fitting results for mock galaxy MG2. For details, see the description of Fig.~\ref{fig:corner1}.}
\label{fig:mg2}
\end{figure*}

\begin{figure*}
\centering
\includegraphics[scale=0.3]{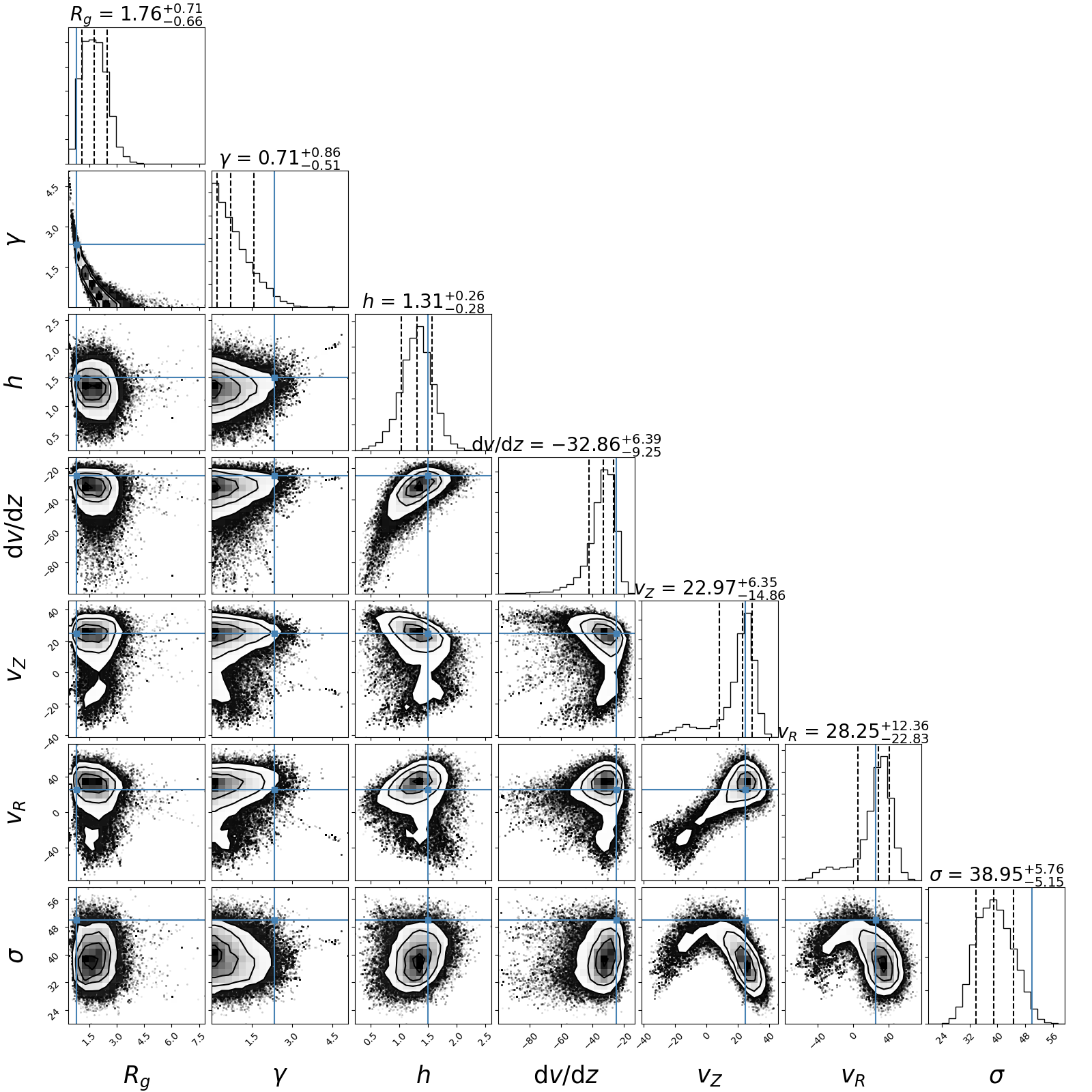}
\includegraphics[scale=0.3]{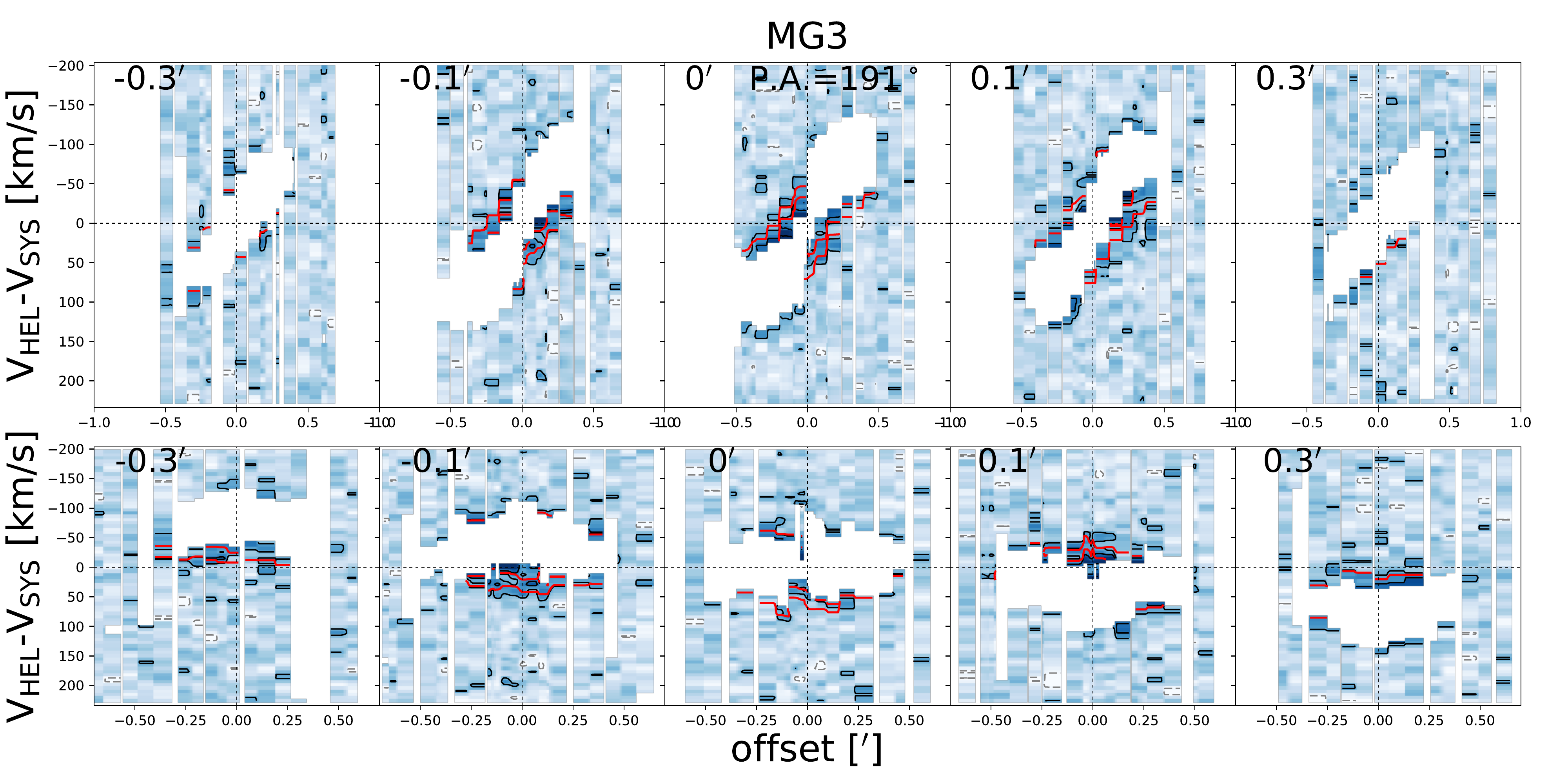}
\caption{MCMC fitting results for mock galaxy MG3. For details, see the description of Fig.~\ref{fig:corner1}.}
\label{fig:mg3}
\end{figure*}

\begin{figure*}
\centering
\includegraphics[scale=0.3]{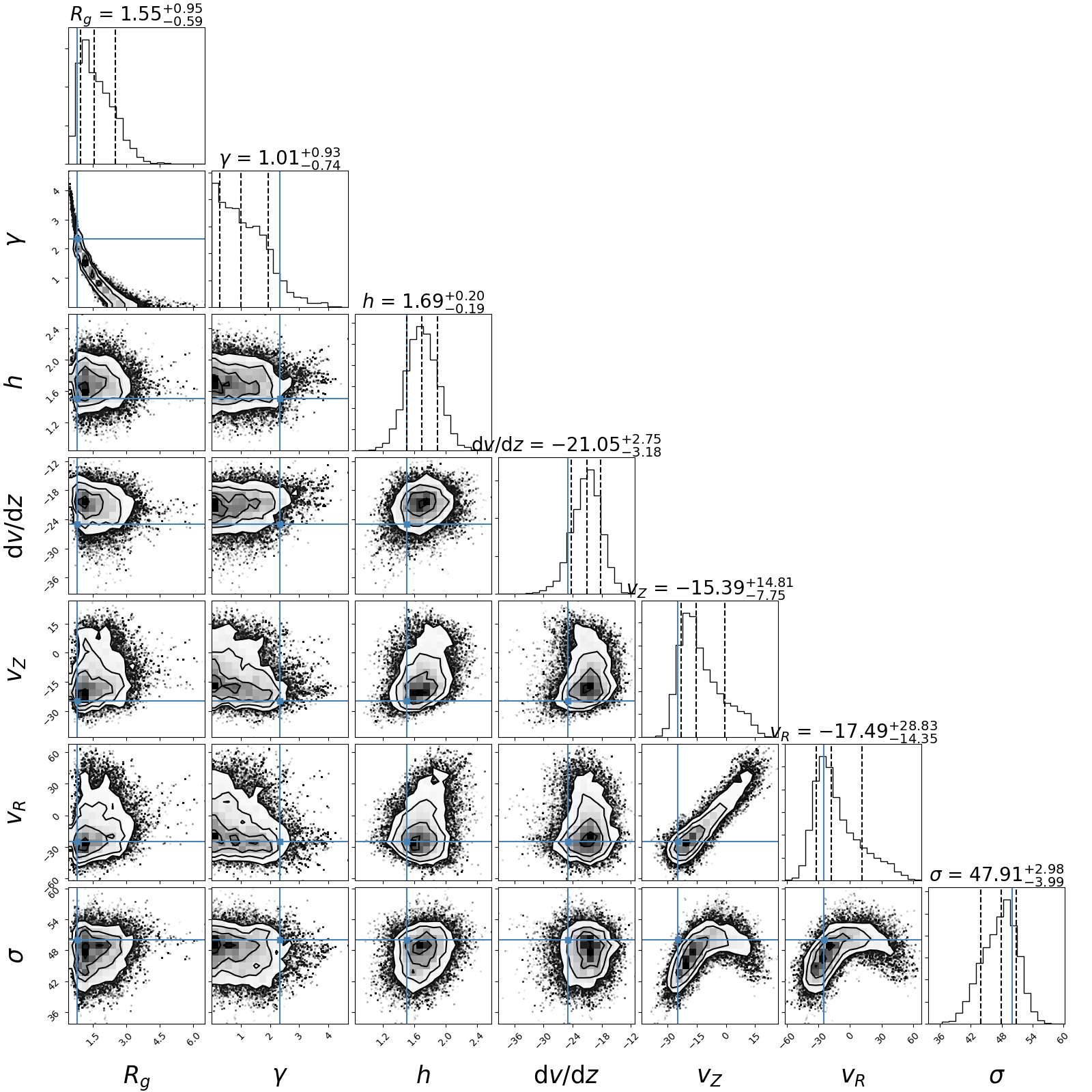}
\includegraphics[scale=0.3]{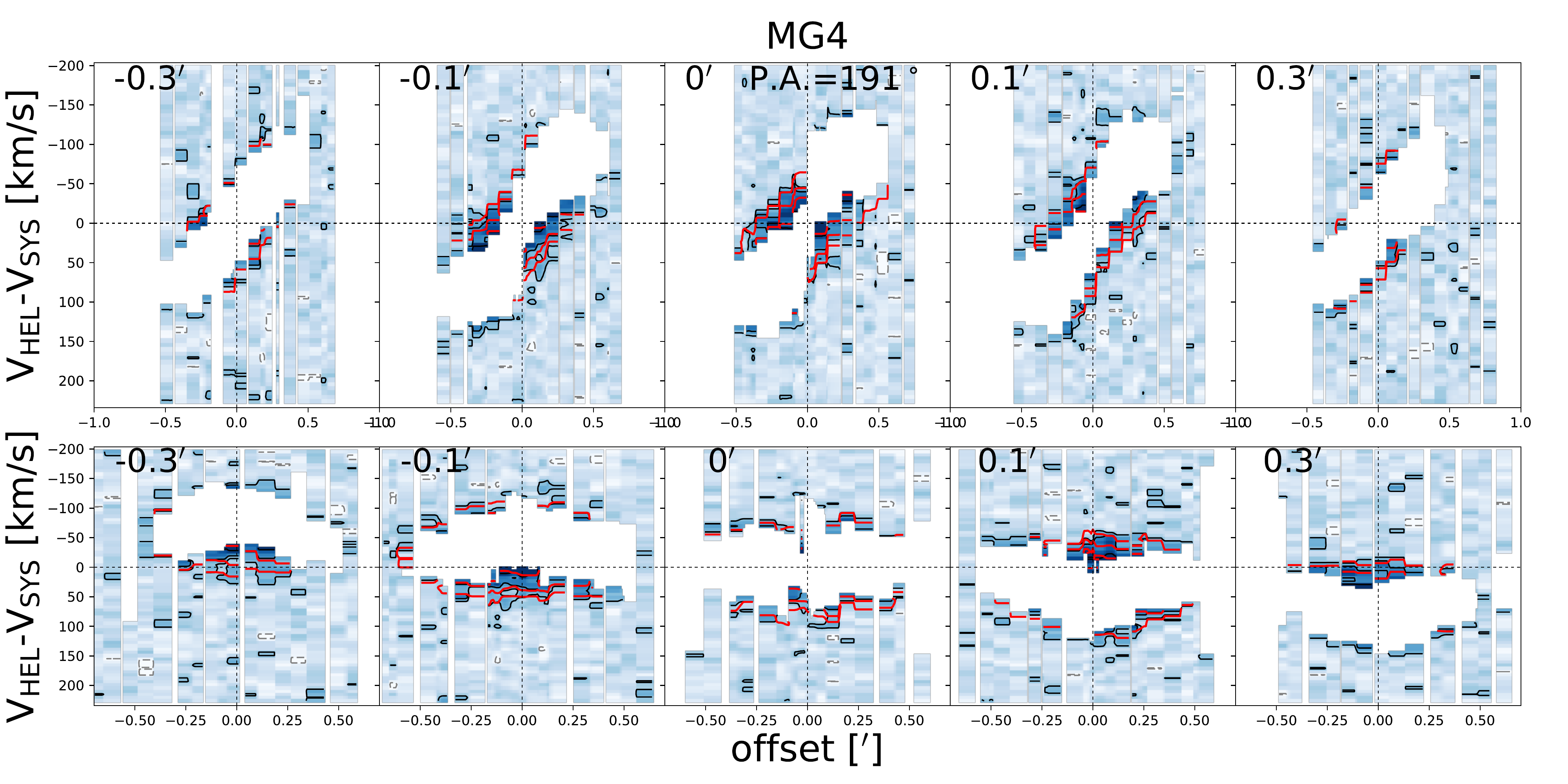}
\caption{MCMC fitting results for mock galaxy MG4. For details, see the description of Figure~\ref{fig:corner1}.}
\label{fig:mg4}
\end{figure*}

\begin{figure*}
\centering
\includegraphics[scale=0.3]{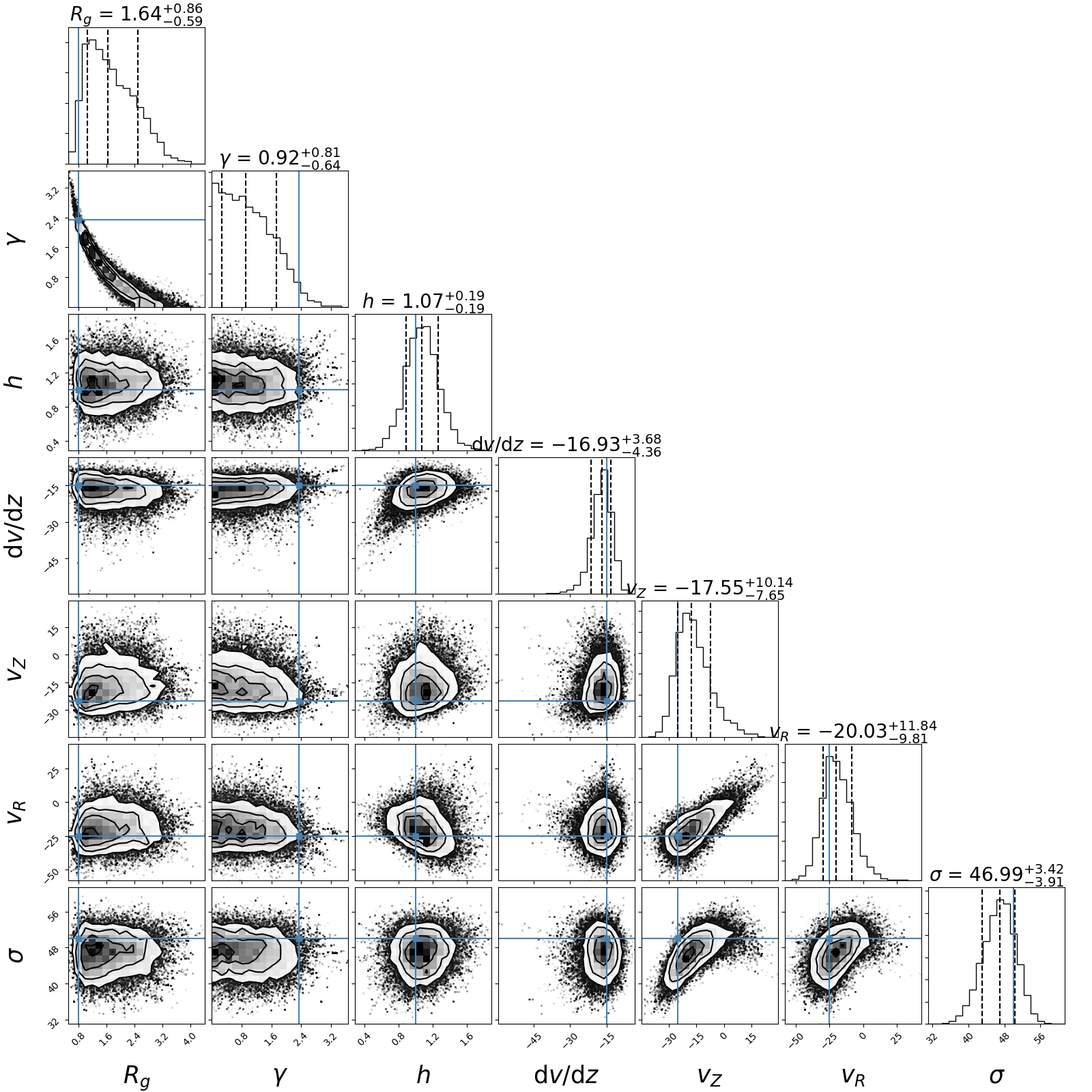}
\includegraphics[scale=0.3]{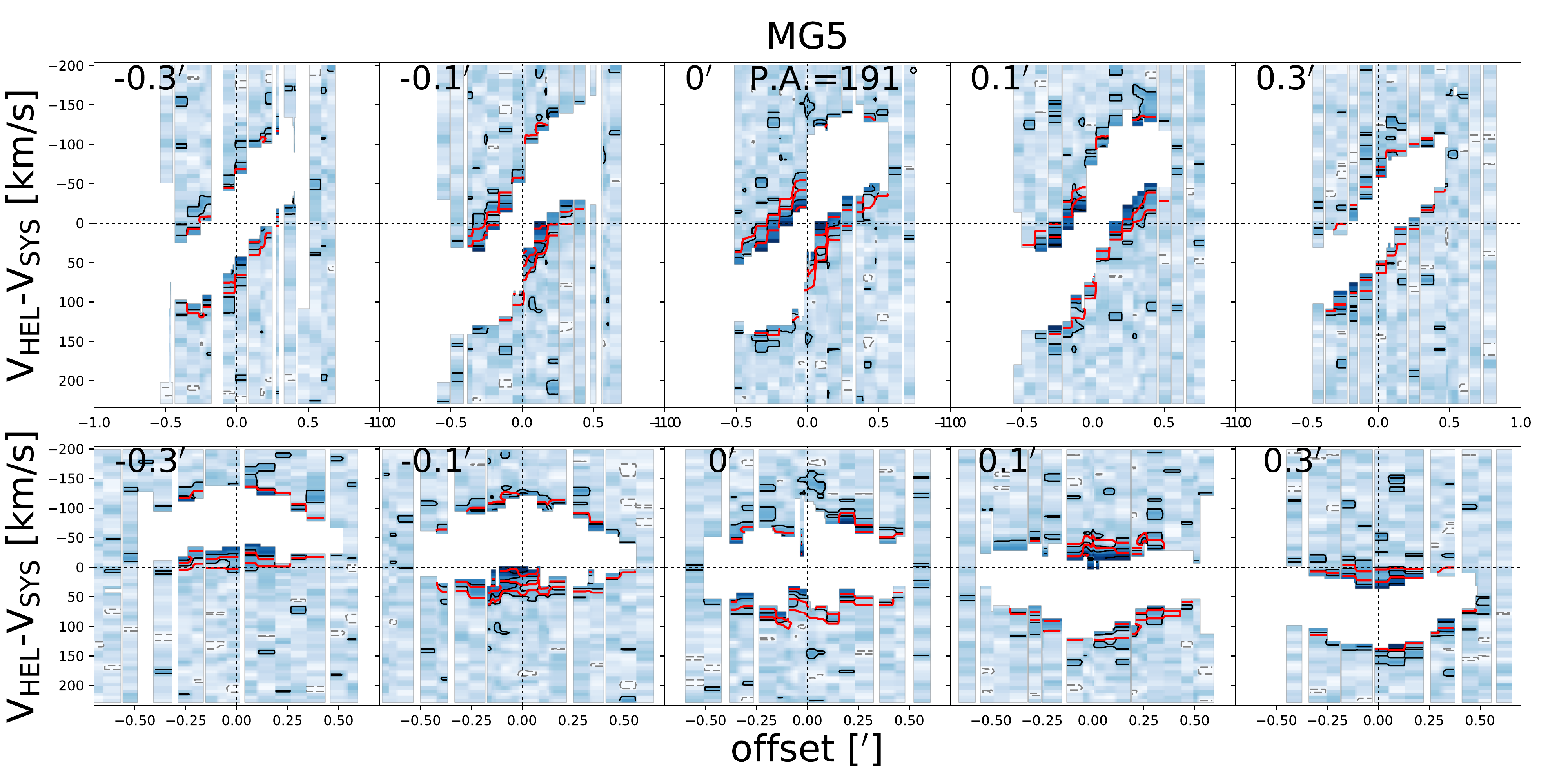}
\caption{MCMC fitting results for mock galaxy MG5. For details, see the description of Fig.~\ref{fig:corner1}.}
\label{fig:mg5}
\end{figure*}

\begin{figure*}
\centering
\includegraphics[scale=0.3]{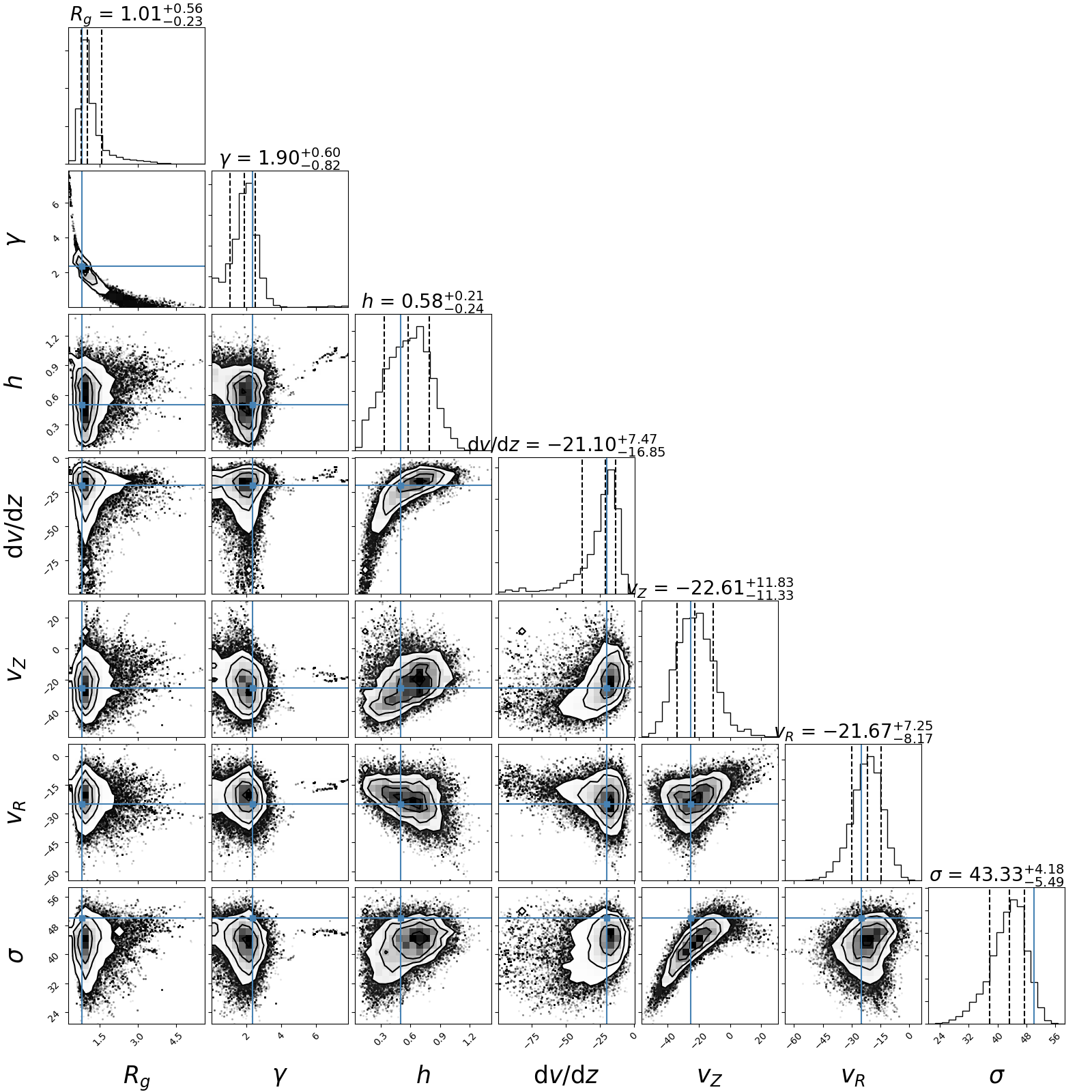}
\includegraphics[scale=0.3]{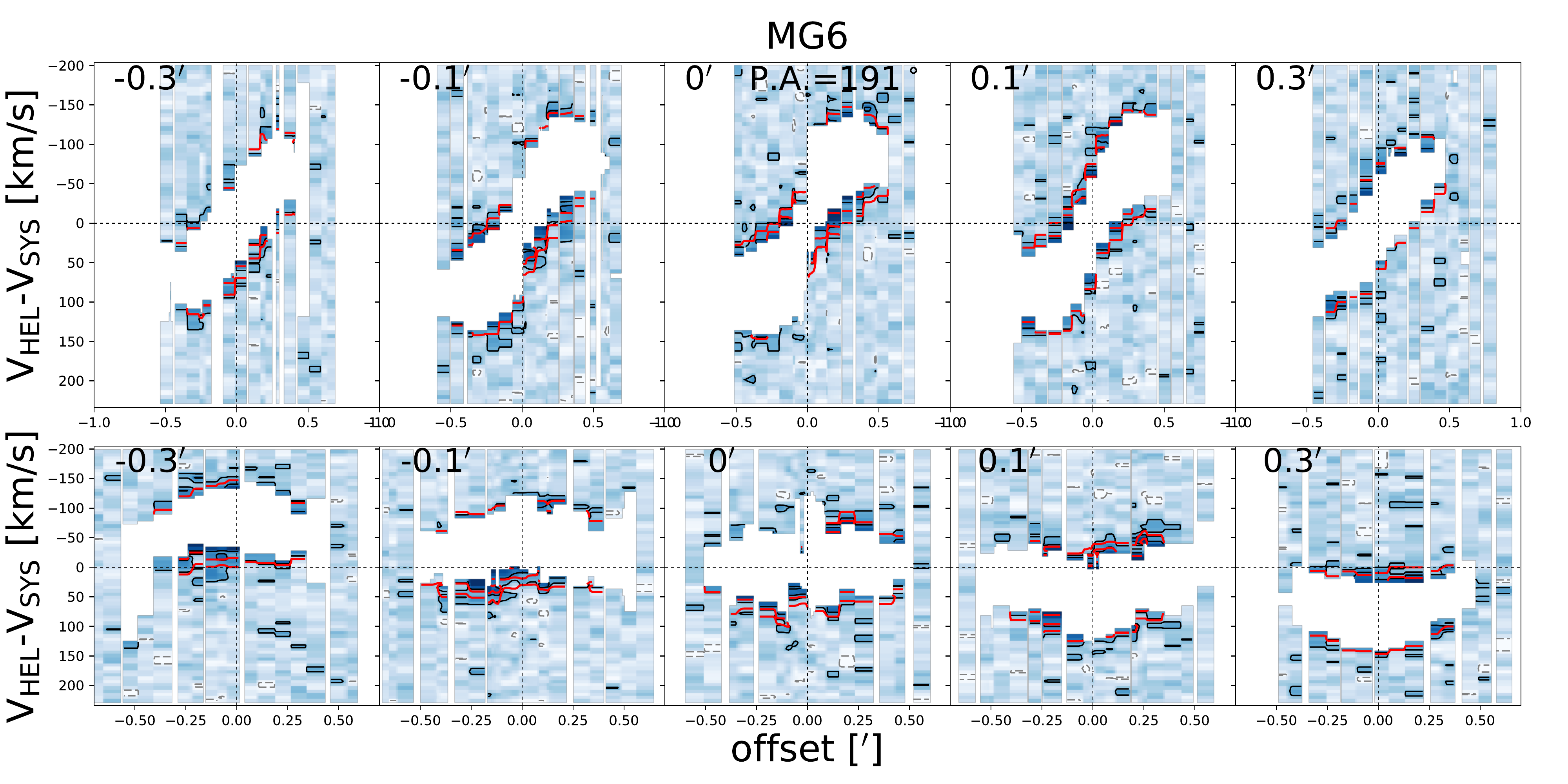}
\caption{MCMC fitting results for mock galaxy MG6. For details, see the description of Fig.~\ref{fig:corner1}.}
\label{fig:mg6}
\end{figure*}

\begin{figure*}
\centering
\includegraphics[scale=0.3]{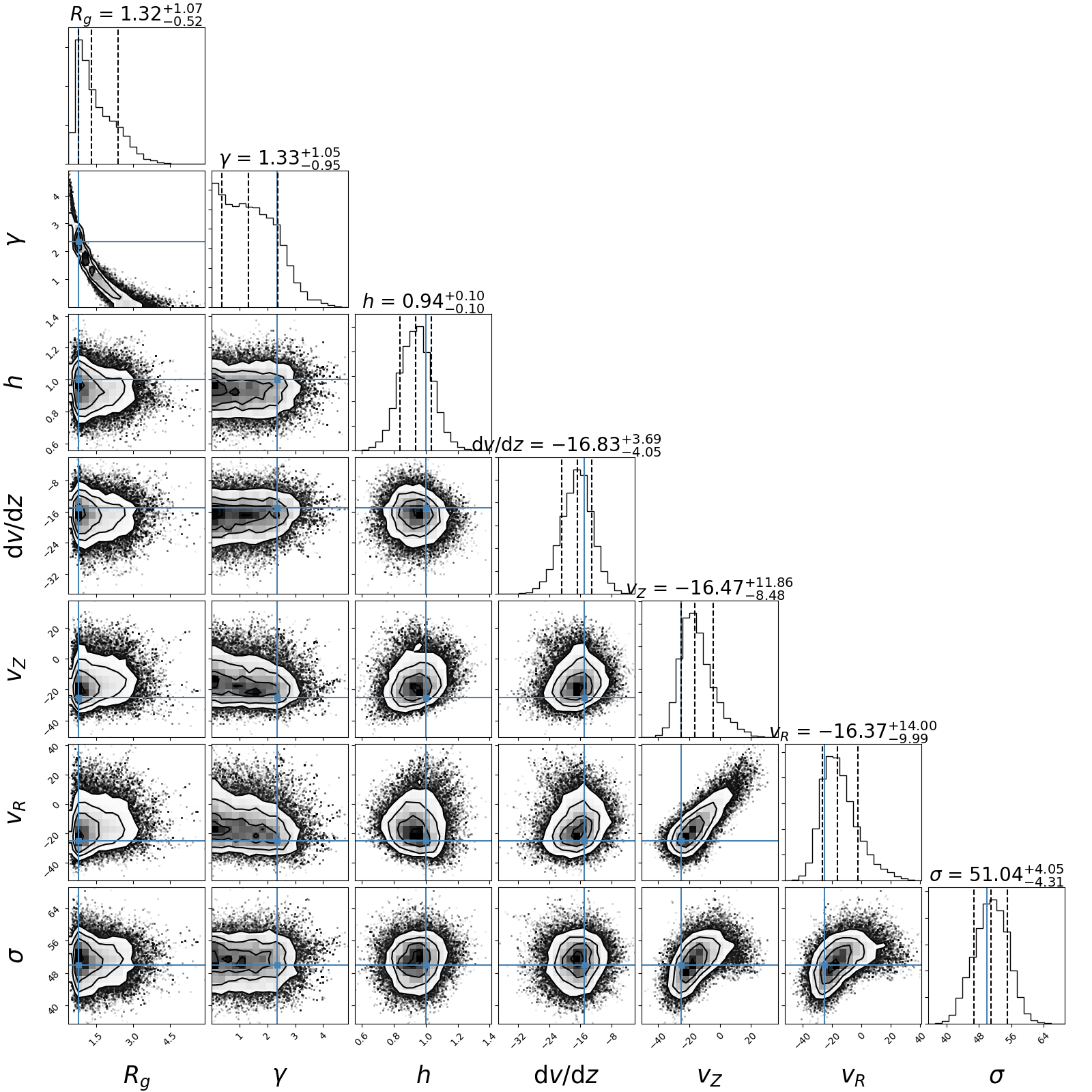}
\includegraphics[scale=0.3]{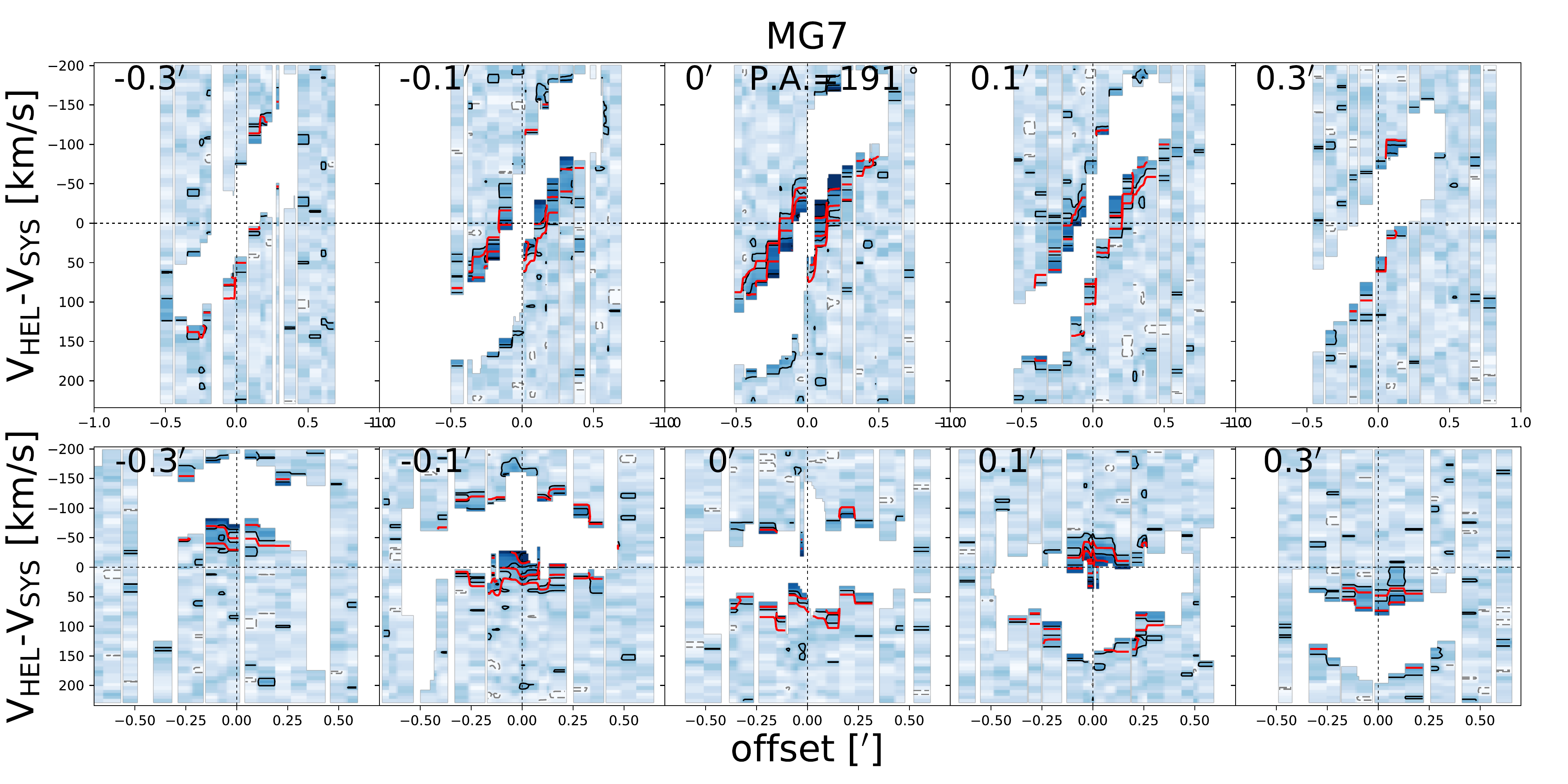}
\caption{MCMC fitting results for mock galaxy MG7. For details, see the description of Fig.~\ref{fig:corner1}.}
\label{fig:mg7}
\end{figure*}

\begin{figure*}
\centering
\includegraphics[scale=0.3]{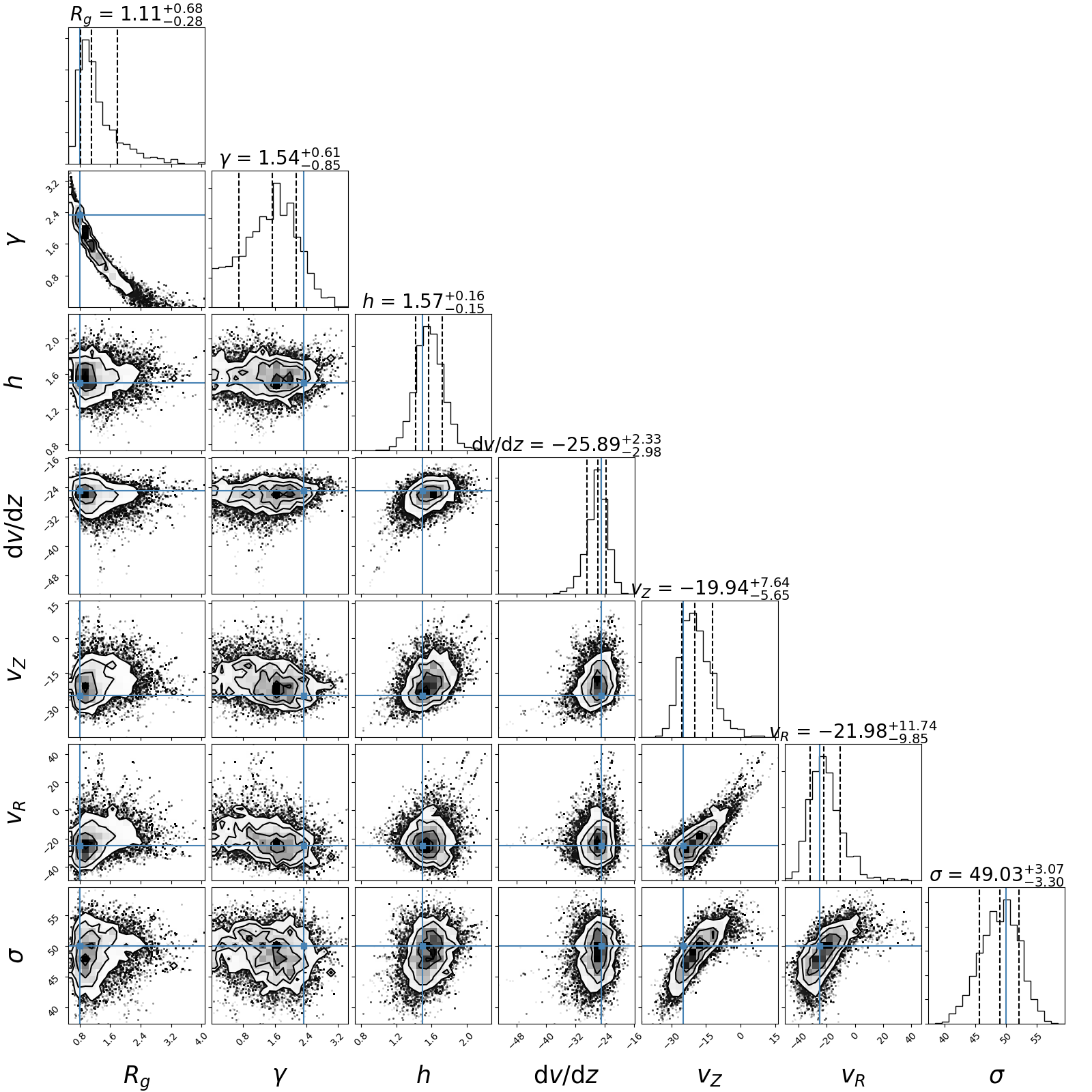}
\includegraphics[scale=0.3]{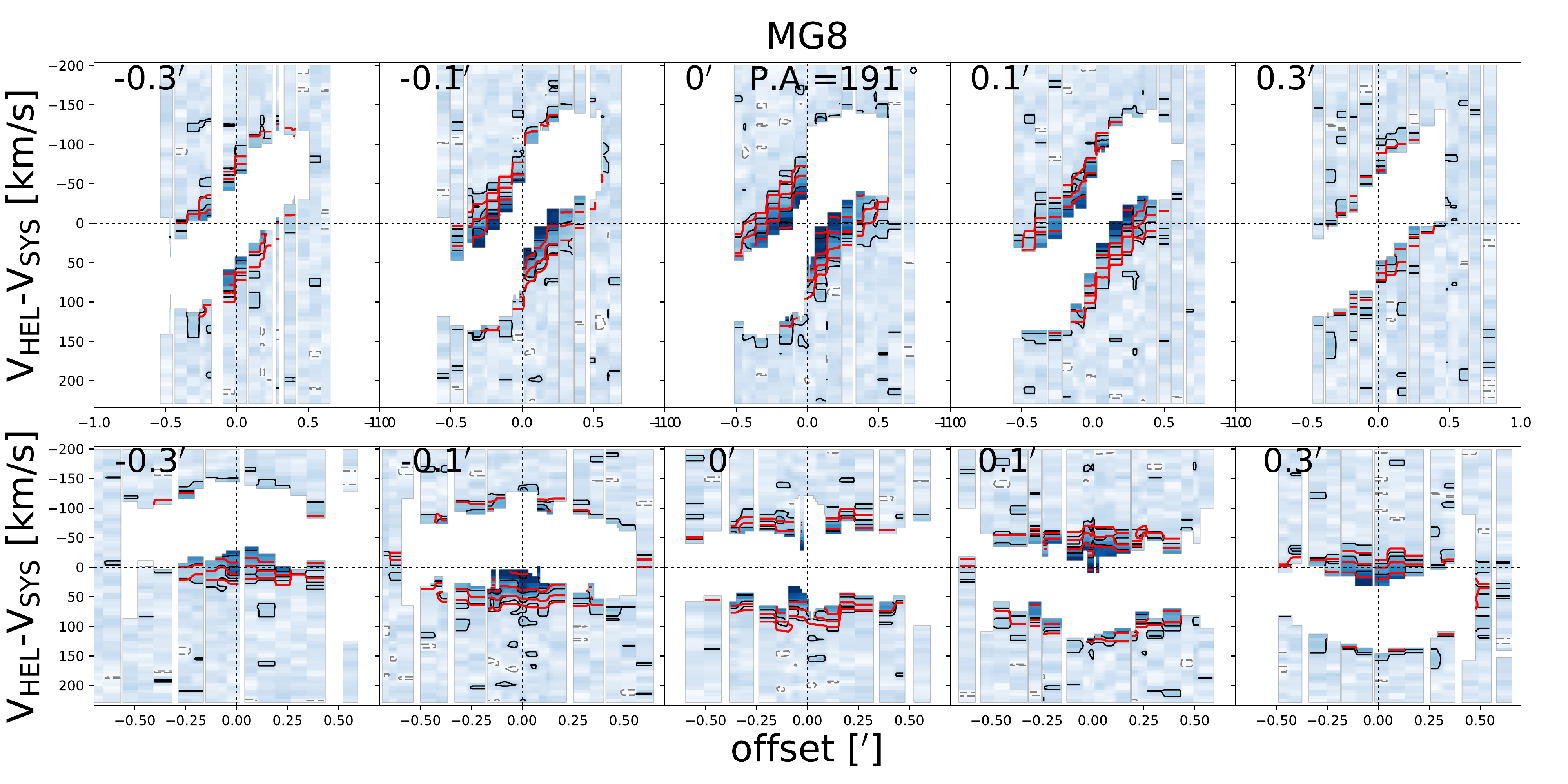}
\caption{MCMC fitting results for mock galaxy MG8. For details, see the description of Fig.~\ref{fig:corner1}.}
\label{fig:mg8}
\end{figure*}

\begin{figure*}
\centering
\includegraphics[scale=0.3]{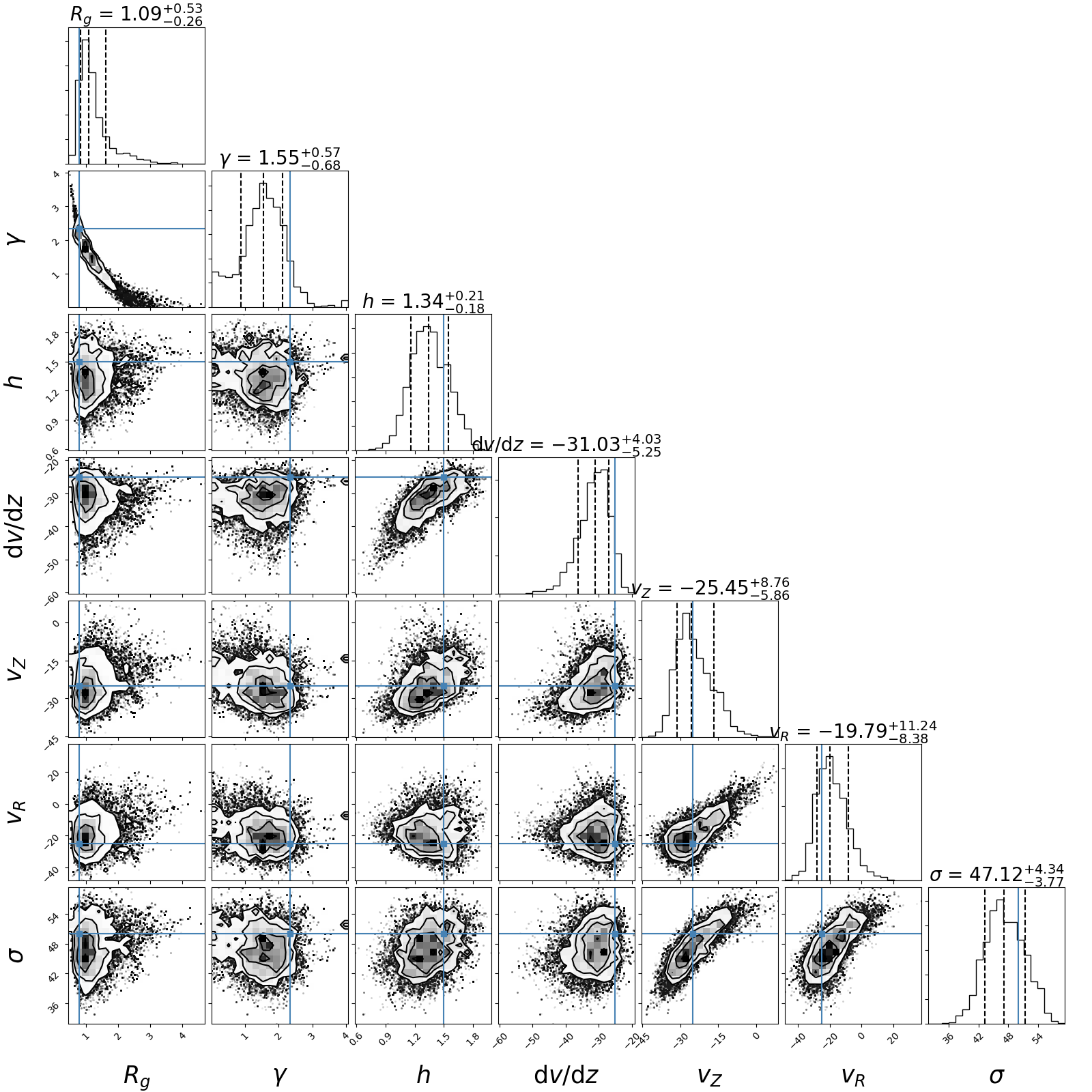}
\includegraphics[scale=0.3]{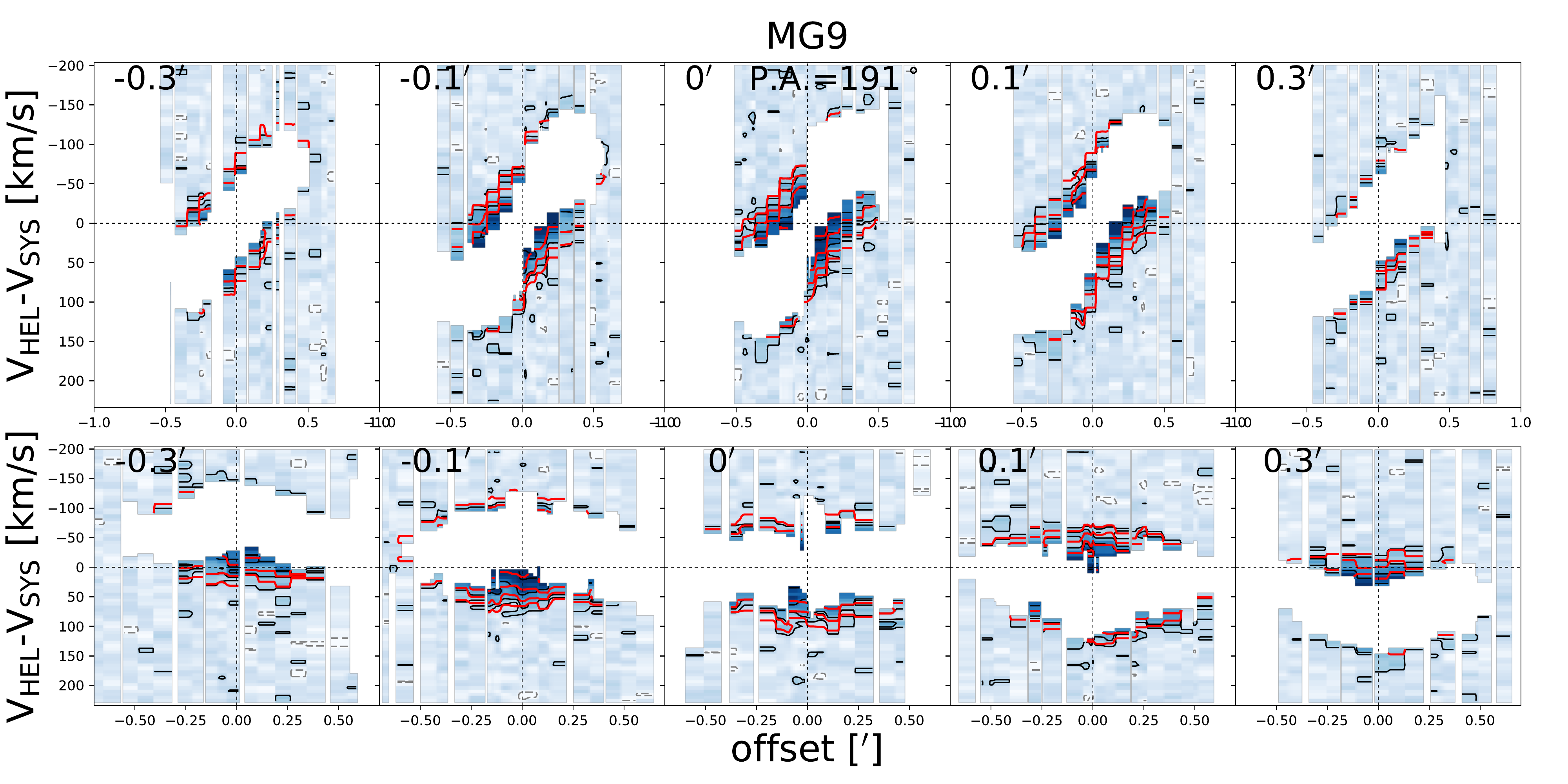}
\caption{MCMC fitting results for mock galaxy MG9. For details, see the description of Fig.~\ref{fig:corner1}.}
\label{fig:mg9}
\end{figure*}

\begin{figure*}
\centering
\includegraphics[scale=0.3]{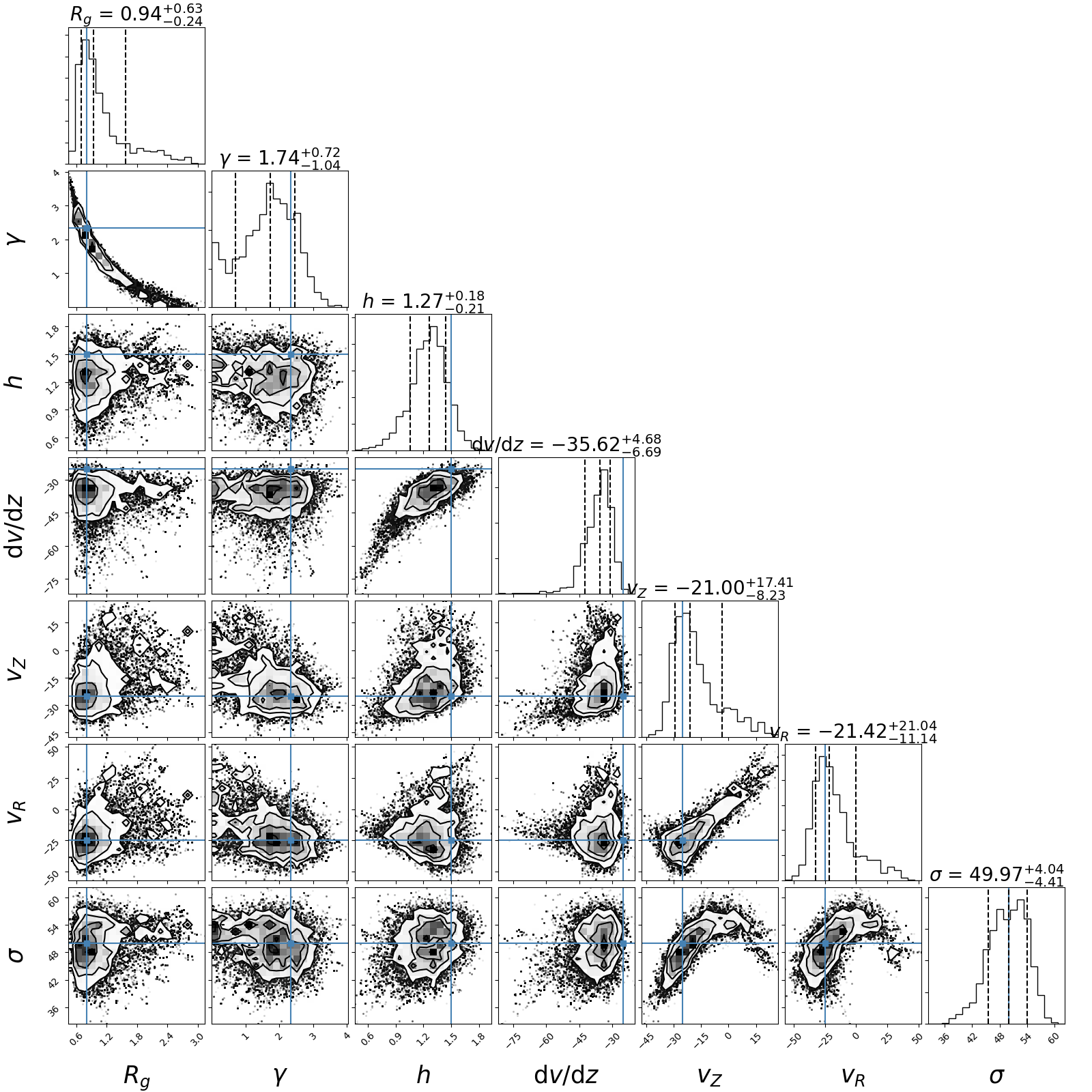}
\includegraphics[scale=0.3]{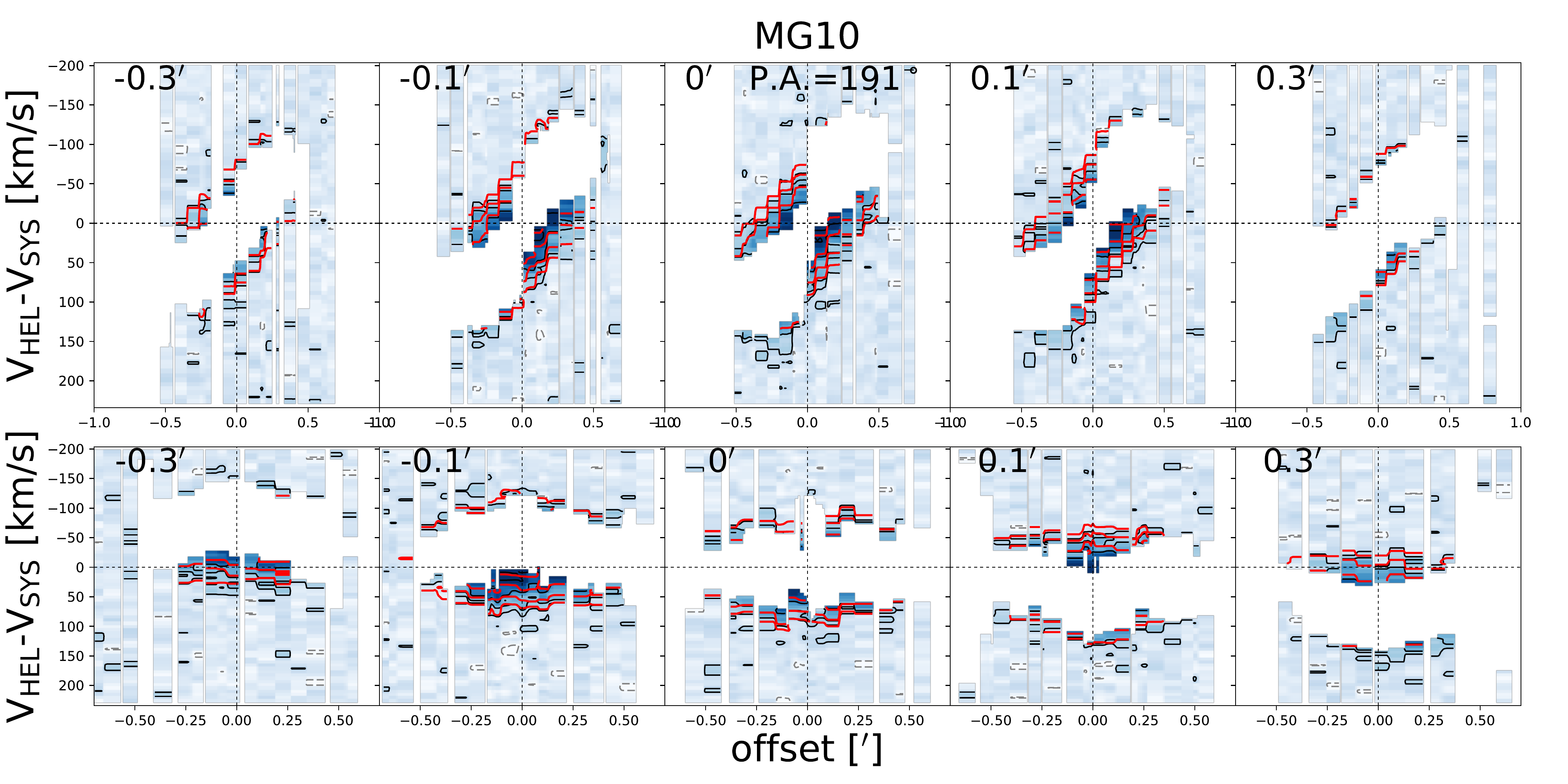}
\caption{MCMC fitting results for mock galaxy MG10. For details, see the description of Fig.~\ref{fig:corner1}.}
\label{fig:mg10}
\end{figure*}

\begin{figure*}
\centering
\includegraphics[scale=0.3]{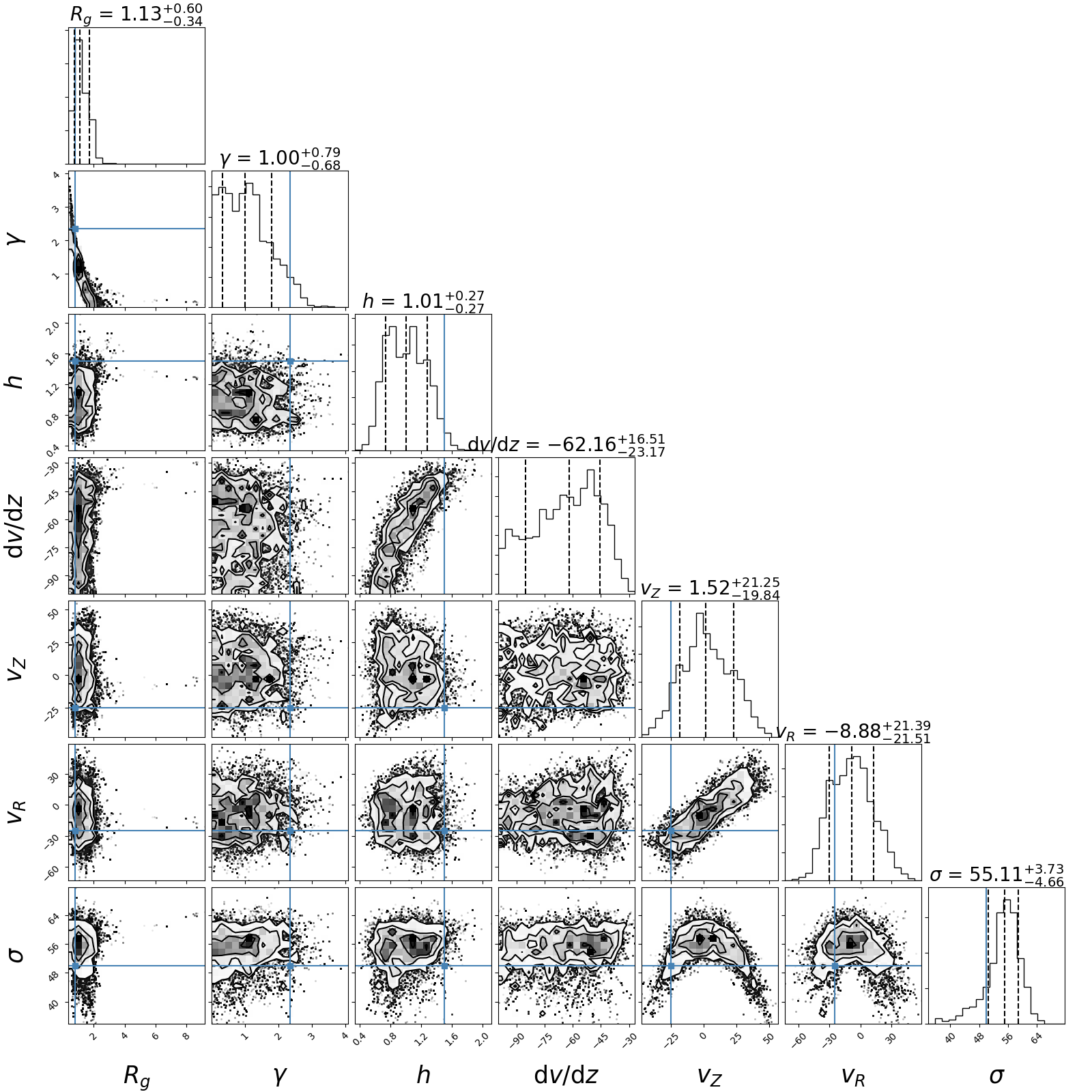}
\includegraphics[scale=0.3]{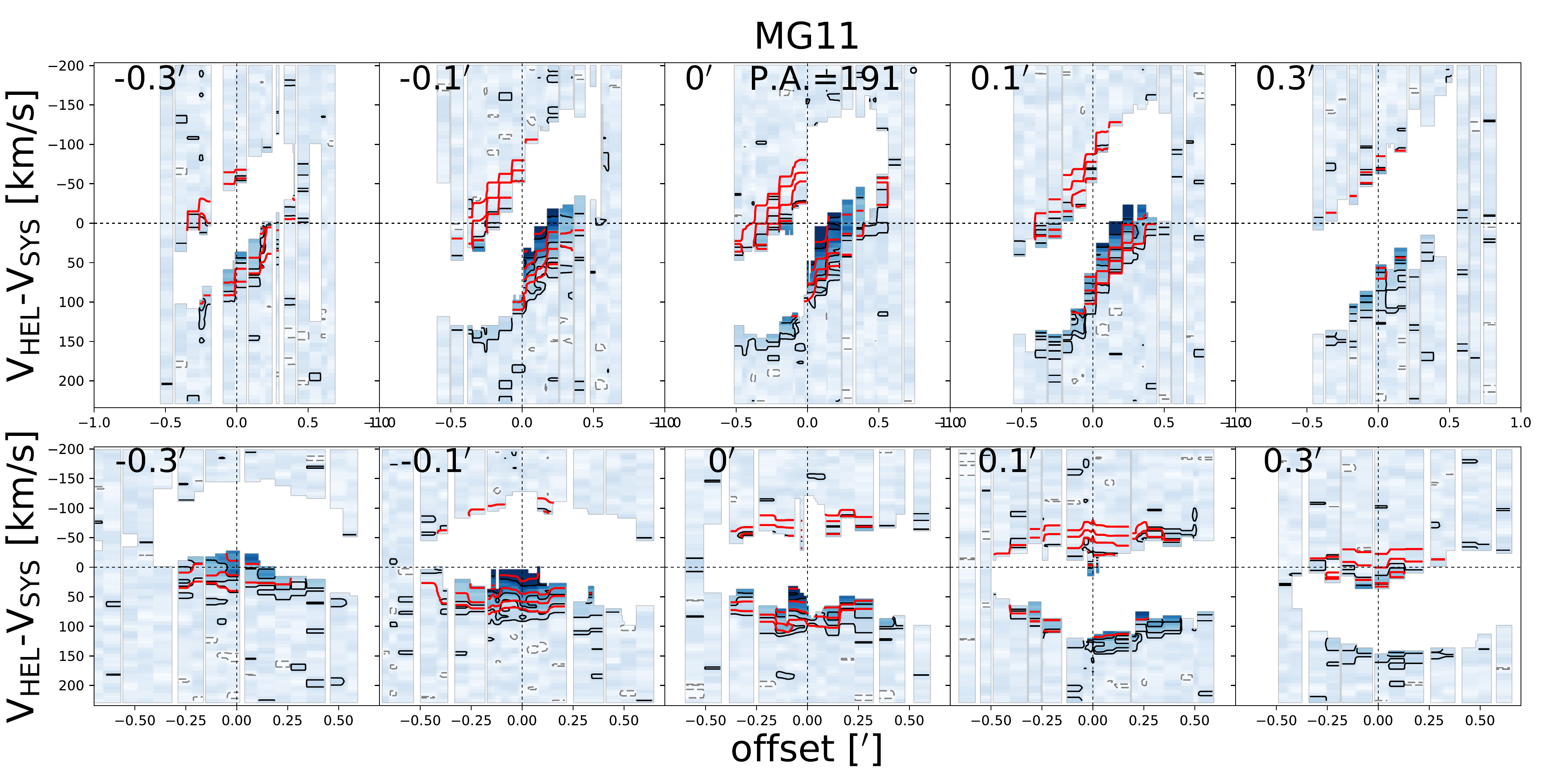}
\caption{MCMC fitting results for mock galaxy MG11. For details, see the description of Fig.~\ref{fig:corner1}.}
\label{fig:mg11}
\end{figure*}

\label{lastpage}

\clearpage
\end{document}